\documentclass[11pt]{article}

\usepackage{amssymb}
\usepackage{amsfonts}
\usepackage{amsmath}
\usepackage{bm}
\usepackage{bbm}
\usepackage{graphicx}
\usepackage{xcolor}
\usepackage{dsfont}
\usepackage{multirow}
\usepackage{hyperref} 

\usepackage{cite}
\allowdisplaybreaks[1]

  \def\pd{\partial} 

  \newcommand{\sumint}[1]{{\hbox{$\textstyle\sum$}\!\!\!\!\!\!\!\int\,}_{\!\!\!\!\raise-0.5ex\hbox{$\scriptstyle{#1}$}}}

\newcommand{\dd}{{\rm d}}
\newcommand{\ie}{{i.e.}}  
\newcommand{\eg}{{e.g.}}

\newcommand{\la}[1]{\label{#1}}
\newcommand{\be}{\begin{equation}}
\newcommand{\ee}{\end{equation}}
\newcommand{\bea}{\begin{eqnarray}}
\newcommand{\eea}{\end{eqnarray}}
\newcommand{\rmi}[1]{{\mbox{\scriptsize #1}}}
\newcommand{\rmii}[1]{{\mbox{\tiny\rm{#1}}}}
\newcommand{\rmiii}[1]{{\mbox{\tiny{$\scriptstyle{\rm#1}$}}}}

\newcommand{\eq}{Eq.~}
\newcommand{\eqs}{Eqs.~}

\newcommand{\nr}[1]{(\ref{#1})}
\newcommand{\tr}{{\rm Tr\,}}
\newcommand{\nn}{\nonumber \\}

\renewcommand{\eq}{eq.~}
\renewcommand{\eqs}{eqs.~}

\newcommand{\mD}{m_\rmii{D}}

\newcommand{\eps}{\epsilon}
\newcommand{\nf}{N^{ }_\rmi{f}}
\newcommand{\Nc}{N_{\rm c}}

\newcommand{\gammaE}{\gamma_\rmii{E}}
\newcommand{\rmO}{{\mathcal{O}}}

\newcommand{\CF}{C_\rmii{F}}

\def\lsi{\raise0.3ex\hbox{$<$\kern-0.75em\raise-1.1ex\hbox{$\sim$}}}
\def\gsi{\raise0.3ex\hbox{$>$\kern-0.75em\raise-1.1ex\hbox{$\sim$}}}

\newcommand{\lsim}{\mathop{\lsi}}
\newcommand{\gsim}{\mathop{\gsi}}

\newcommand{\nF}{n_\rmii{F}}
\newcommand{\nB}{n_\rmii{B}}
\newcommand{\re}{\mathop{\mbox{Re}}}
\newcommand{\im}{\mathop{\mbox{Im}}}
\newcommand{\Tint}[1]{{\hbox{$\sum$}\!\!\!\!\!\!\!\int\,}_{\!\!\!\!\raise-0.9ex\hbox{$\scriptstyle{#1}$}}}
\newcommand{\Tinti}[1]{{{\Sigma}\!\!\!\!\raise0.3ex\hbox{$\int$}_\rmii{${#1}$}}}
\newcommand{\unit}{{\mathbbm{1}}} 
\newcommand{\bi}{\begin{itemize}}
\newcommand{\ei}{\end{itemize}}
\newcommand{\hide}[1]{ }
\newcommand{\bsl}[1]{\,\slash\!\!\!\!{#1}\,}

\newcommand{\deltabar}{\raise-0.02em\hbox{$\bar{}$}\hspace*{-0.8mm}{\delta}}
\newcommand{\ddeltabar}{\raise-0.18em\hbox{$\bar{}$}\hspace*{-0.8mm}{\delta}}
\newcommand{\Rate}{\Gamma}  
\renewcommand{\P}{\mathcal{P}}

\newcommand{\K}{\mathcal{K}}
\newcommand{\Q}{\mathcal{Q}}
\newcommand{\X}{\mathcal{X}}

\newcommand{\lif}{l^{ }_\rmi{2f}}
\newcommand{\ltf}{l^{ }_\rmi{3f}}

\newcommand{\lib}{l^{ }_\rmi{2b}}
\newcommand{\ltb}{l^{ }_\rmi{3b}}
\newcommand{\fin}{\mbox{\sl f\,}}
\newcommand{\ini}{\mbox{\sl i\,}}
\newcommand{\sfin}{\rmii{\sl f\,}}
\newcommand{\sini}{\rmii{\sl i\,}}

\newcommand{\Sfunc}{\mathcal{F}}

\renewcommand{\vec}{\bm}
\renewcommand{\ini}{{\textsl{i}}} 
\renewcommand{\fin}{{\textsl{f}}} 

\makeatletter \@addtoreset{equation}{section} \makeatother
\renewcommand{\theequation}{\arabic{section}.\arabic{equation}}
\makeatletter
\renewcommand\section{\@startsection {section}{1}{\z@}%
                                   {-5.5ex \@plus -1ex \@minus -.2ex}
                                   {2.3ex \@plus.2ex}%
                                   {\normalfont\large\bfseries}}
\renewcommand\subsection{\@startsection{subsection}{2}{\z@}%
                                     {-3.25ex\@plus -1ex \@minus -.2ex}%
                                     {1.5ex \@plus .2ex}%
                                     {\normalfont\normalsize\bfseries}}
\renewcommand\thesection {\@arabic\c@section}
\renewcommand\thesubsection   {\thesection.\@arabic\c@subsection}
\renewcommand{\@seccntformat}[1]{%
\csname the#1\endcsname.\hspace{1.0em}}
\makeatother

\begin{document}

\begin{titlepage}

\begin{flushright}
November 2026 
\end{flushright}

\begin{centering}
\vfill

{\Large{\bf
  Collisional energy loss distribution of a fast parton \\[3mm]
  in a hot or dense QCD medium }}

\vspace{0.8cm}

G.~Jackson
and 
S.~Peign\'e

\vspace{0.8cm}

{\em
SUBATECH 
  (IMT Atlantique, 
Nantes Universit\'e, 
  IN2P3/CNRS), \\ 
4 rue Alfred Kastler, 
La Chantrerie BP 20722, 
44307 Nantes, France \\}

\vspace*{0.6cm}

{\em 
Emails: 
  \href{mailto:jackson@subatech.in2p3.fr}{jackson@subatech.in2p3.fr}, 
  \href{mailto:peigne@subatech.in2p3.fr}{peigne@subatech.in2p3.fr}
}

\vspace*{0.8cm}

\mbox{\bf Abstract}
 
\end{centering}

\vspace*{0.3cm}
 
\noindent
We compute the probability distribution for collisional energy loss of 
an ultrarelativistic parton crossing a quark-gluon plasma. 
This {\em collisional quenching weight} 
has not been determined previously, unlike 
the average collisional loss per unit distance, 
although it should be a more accurate quantity to use in 
jet-quenching phenomenology. 
The quenching weight is obtained from a well-known kinetic equation 
which resums an arbitrary number of elastic scatterings of 
the energetic parton with the medium, providing a complete description 
of the stochastic energy exchange, 
including the possibility of energy gain from thermal fluctuations. 
The formulation also naturally extends the standard treatment of 
collisional energy loss to finite path lengths, 
which could be relevant not only for heavy-ion collisions, 
but also for light-ion, and possibly 
proton-nucleus and proton-proton collisions. 
We predict the quenching weight in a setup where individual 
elastic scatterings are described using the 
hard thermal loop approximation for soft exchanges, 
with a smooth matching to the hard domain. 

\vfill

\end{titlepage}

\tableofcontents

%
\section{Introduction}
\label{se:intro}

In a famous paper~\cite{Bjorken:1982tu}, 
Bjorken addressed the collisional energy loss 
suffered by energetic light partons going 
through a quark-gluon plasma (QGP) produced in 
pp collisions. 
He suggested that parton energy loss could lead to 
suppression of high-$p_\rmii{T}$ jets in hadron-hadron collisions, 
an effect now called {\em jet-quenching}. 
As a signal for QGP formation,\footnote{%
  The effect, 
  which generically includes also the  
  nuclear suppression of high-$p_\rmii{T}$ single hadron production 
  in heavy ion collisions, 
  has been first well-established by 
  the RHIC experiments~\cite{PHENIX:2001hpc,PHENIX:2003qdj,STAR:2002ggv,PHENIX:2005nhb,STAR:2006btx}.
} 
jet-quenching 
has driven numerous studies over the following decades, 
both experimental and 
theoretical~\cite{dEnterria:2009xfs,Majumder:2010qh,Cao:2020wlm}. 
Parton energy loss has been confirmed as a 
crucial nuclear 
effect in heavy ion collisions, and is generally considered to 
result from two contributions: 
collisional (\ie \ occurring in elastic collisions) and radiative 
(gluon bremsstrahlung)~\cite{Qin:2007rn}. 
The radiative part is often the main one, e.g.\ 
for the energy loss of light partons crossing a 
QGP of sufficiently large size $L\,$.
In general however, collisional energy loss may also be 
sizeable and play a role in phenomenology. 
This seems to be the case for heavy quarks~\cite{Rapp:2009my}, 
and for QGPs of ``small'' size~\cite{Faraday:2025pto}.
Indeed, measurements in high-multiplicity pp and pPb collisions 
(as well as peripheral AA collisions) 
continue to motivate discussions about whether medium-like effects 
can exist in small systems~\cite{CMS:2016fnw,ATLAS:2016yzd,ATLAS:2017rtr}. 
%
%
Recent experimental results for 
$R_\rmii{OO}$~\cite{CMS:2025bta,ALICE:2026zck}
have been compared to various theoretical predictions 
using different model assumptions. 
Some of these theoretical studies include the effect 
of collisional energy loss on the light-ion 
data
~\cite{Zakharov:2021uza,Faraday:2025pto,Faraday:2025prr,Hulcher:2026dht,Kudinoor:2026wcs,Ke:2022gkq,Vitev:2026phi}. 
%
Note that the possible relevance of collisional loss for 
heavy quarks and/or small $L$ is suggested by the parametric dependence 
of the average radiative and collisional losses on the various 
parameters of the problem, see ref.~\cite{Peigne:2008wu}.

There are therefore good motivations to study 
the collisional energy loss of a fast parton crossing a QGP. 
However, 
calculations of in-medium collisional energy loss 
have focused mostly on the average energy loss
$\Delta E_{\rm coll}$ in a medium of size $L\,$, namely,  
$\Delta E_{\rm coll} = ({\rm d}E_\rmi{coll}/{\rm d}x) \cdot L\,$, 
where ${\rm d}E_\rmi{coll}/{\rm d}x$
is the  average loss per unit distance. 
Bjorken's initial leading-logarithm (LL) calculation of 
${\rm d}E_\rmi{coll}/{\rm d}x$~\cite{Bjorken:1982tu} was 
improved upon in ref.~\cite{Thoma:1990fm}, where 
Hard Thermal Loop (HTL) resummation was implemented. 
The calculation of ${\rm d}E_\rmi{coll}/{\rm d}x$ was later revisited, 
for a heavy fermion in a hot QED plasma to 
next-to-leading logarithmic (NLL) 
accuracy~\cite{Braaten:1991jj,Peigne:2007sd}, and 
for an energetic parton in a QGP with running coupling effects taken 
into account (to LL in ref.~\cite{Peshier:2006hi} 
and to NLL in ref.~\cite{Peigne:2008nd}). 
While higher-order corrections are important, the average parton energy loss 
is likely inadequate for realistic applications. 

A more useful quantity would be the actual probability distribution, 
or ``quenching weight'',  needed for instance to evaluate $R_\rmii{AA}$ 
(obtained by a convolution with production cross sections). 
Quenching weights associated with radiative parton energy loss 
(encoding the fluctuating number and energies of emitted gluons) 
have been studied in several places~\cite{Baier:2001yt,Salgado:2003gb}, 
and have been deployed in studies of 
small systems~\cite{Huss:2020whe,Huss:2020dwe}. 
%
%
As for the energy loss accumulated in multiple elastic collisions, 
the associated quenching weight has received less attention.
In ref.~\cite{Peshier:2006mp}, 
the collisional energy loss probability distribution was derived 
for a {\em single} scattering. 
In ref.~\cite{Faraday:2024gzx}, 
some models are proposed, but apparently no derivation from first-principles.  
Thus, a derivation of the collisional loss distribution in a QCD medium would 
fill a conceptual gap in the modelling of 
jet/hadron and heavy flavour suppression. 
As mentioned above, there are indications  
that collisional loss can play a role at small $L$. 
The probability distribution should be all the more useful since, 
for small values of $L$, 
significant fluctuations in energy loss are expected, 
which are not taken into account by the average energy loss 
(i.e., by the first moment of the distribution). 

The problem of deriving the energy loss distribution 
of a fast charge in matter is, of course, not new. 
The stochastic nature of energy loss in this situation 
may be described with a probability distribution, 
first calculated by Landau in 1944~\cite{Landau:1944fvs}. 
He recognised that large energy transfers 
suffered by a fast charge crossing a medium 
of finite length 
are dominated by rare but very hard collisions on atoms, 
arising from the tail of the ionisation spectrum, 
thus contributing sizable event-by-event fluctuations 
about the most probable energy loss. 
This phenomenon, known as {\em straggling}, 
is captured by a skewed distribution 
whose higher moments all diverge due to a non-Gaussian 
behaviour $\sim 1/\Delta^2$ at large energy transfers $\Delta\,$. 
It plays a role 
in modern particle detector physics,\footnote{%
  See, for example, sec.~34.2.9 of the 
  2024 Review of Particle Physics~\cite{ParticleDataGroup:2024cfk}. 
} 
and has been the subject of theoretical refinements 
and measurements over the 
years~\cite{Bak:1987cz,Bichsel:1988if,Trofymenko:2022cdf}. 
The energy transfer spectrum for a single Coulomb scattering 
is responsible for this power-law tail, and 
an entirely analogous situation in QCD arises from 
scatterings involving $t$-channel gluon exchange. 
Consequently, the central limit theorem  is not automatic 
and convergence to a normal distribution 
may be parametrically slow when rare hard exchanges contribute significantly 
to the total variance~\cite{lindeberg1922}. 
In heavy-ion physics, near-thermal heavy quarks 
are often treated with diffusion 
or Langevin approaches~\cite{Moore:2004tg,Ghiglieri:2015ala}, 
but for highly energetic partons traversing a medium of finite size, 
fluctuations can push one far beyond the Gaussian regime usually assumed. 
In this paper we derive the {\it collisional quenching weight} 
$f(x, \Delta)\,$, i.e. the probability distribution for the 
collisional energy loss $\Delta\,$,
of a fast charge crossing a path length $x$ across 
a hot or dense QCD medium.\footnote{%
  Mind that the path length $x$ may differ 
  from the plasma size $L$, and in particular may be much smaller, 
  when the parton traverses the medium near its periphery.
}

Our work is presented as follows. 
In sec.~\ref{se:setup} we begin with the basic setup, describing 
the kinetic equation for the probability density $f(x, \Delta)$, 
its main theoretical ingredient given by 
the differential single scattering rate $w(\eps)$, 
and its general solution. 
(Elaboration of formal aspects appears in appendix~\ref{app:xsection}.) 
A detailed treatment of $w(\eps)$ appears in 
sec.~\ref{sec:perturbative-rate}, with  
HTL resummation for small $\eps$ 
and exact kinematics in the hard domain. 
(The asymptotic behaviour of the HTL computation is given 
in appendix~\ref{app:Taylor-ex}.) 
Assembling the input from the previous section, 
we solve the kinetic equation in 
sec.~\ref{sec:quenching-weight}, with numerical details in 
appendix~\ref{app:numerics}. 
Finally, we conclude in sec.~\ref{se:disc} 
with a discussion of the implications of our results.

%
\section{Formulation of the problem}
\label{se:setup}

%
\subsection{Kinetic equation and scattering rate}
\label{se:kin-eq}

We want to determine the probability density $f(x,\Delta)$ 
(normalised as $\int {\rm d} \Delta \, f(x,\Delta) = 1$) 
for an incident parton of energy $E_\ini$ 
to undergo energy exchange $\Delta\,$, 
after travelling the distance $x$ inside a QCD medium. 
The shape of this distribution then varies as a function of 
$x$ according to the kinetic equation~\cite{Landau:1944fvs}
\be
  \frac{\pd f}{\pd x} 
  \; = \;
  \int \dd \eps \ 
  w(\eps, E_\ini ) \, 
  \bigl[ 
  f(x,\Delta -\eps) 
  - f(x,\Delta)
  \bigr] 
  \; ,
  \la{eq:kinetic1}
\ee
%
where $w(\eps,E_\ini)$ is the differential interaction rate 
(per unit distance) 
for the incoming parton to ``lose'' the amount $\eps\,$. 
The above equation assumes that the many 
successive energy exchanges incurred by the fast parton in 
elastic collisions happen independently, and that 
the accumulated energy transfer $\Delta$ (and a fortiori the individual 
transfers $\eps$) are much smaller than the projectile 
energy $E_\ini$~\cite{Landau:1944fvs}. 
In the present study we consider a parton 
with asymptotically large energy ($E_\ini \to \infty$), 
and aim to derive $f(x,\Delta)$ for any finite $\Delta\,$. 
We thus automatically have $|\Delta|,|\eps| \ll E_\ini\,$, 
which permits the interaction rate at 
a specific layer depth $x$ 
to be approximated as 
$w\big(\eps, E(x)\big) \simeq w(\eps, E_\ini)$, where 
$E(x)$ would be the particle's actual energy at that moment. 
In other words, 
we do not track the degradation of $E_\ini$ 
needed to understand how the parton 
would eventually reach equilibrium~\cite{Peshier:2008bg,DuPlessis:2026pyr}. 

For our study, the target is a hot and/or dense QCD medium, 
of temperature $T$ and (quark) chemical potential $\mu\,$. 
In the case of a hot medium ($T>0$) in thermal equilibrium, 
there is a non-zero probability for the fast particle to 
receive energy from a thermal parton in the medium. 
In this case, the rate $w(\eps)$ for transitions 
$E_\ini \to E_\fin \, \equiv \, E_\ini - \eps$ is non-zero for $\eps < 0\,$, 
and as a result the support of $f(x,\Delta)$ extends to 
the domain $\Delta < 0\,$, corresponding to {\it energy gain}. 
We stress that the kinetic equation~\nr{eq:kinetic1} also holds 
in presence of energy gain, whereby 
$\int \dd \eps = \int_{-\infty}^{\infty}  \dd \eps$ 
(in the $E_\ini \to \infty$ limit). 
This differs from the situation considered in ref.~\cite{Landau:1944fvs}, 
where the fast particle passes through ordinary matter, 
and where the rate $w(\eps) = n \, \dd \sigma/ \dd \eps$ in 
a single scattering on an atom 
(with $n$ the atomic number density of the medium and 
$\dd \sigma$ the cross section) is non-zero only if $\eps \geq 0\,$, 
accounting for the ionization of target atoms and 
their excitation when bound electrons change energy levels.

Let us gradually introduce the theoretical setup. 
For simplicity, we will take the energetic parton to be 
a heavy quark of mass $M \gg T, \, \mu$ 
(typically a charm or bottom quark). 
Thus the medium contains only gluons and $\nf$ light quark flavours, 
so that the heavy quark can be unambiguously tagged in the final state, 
and its energy loss cleanly defined.\footnote{%
  Considering a heavy quark should be of minor importance to our study. 
  Indeed, when $E_\ini \to \infty$, 
  our results will be independent of the quark mass $M$, 
  and should therefore also apply to high-energy 
  light partons, with a few minor modifications. 
  In particular, 
  the precise definition of energy loss for a light parton 
  has to be adapted due to the presence of exchange diagrams when the 
  parton scatters elastically (via single gluon exchange) 
  on the same type of parton in the plasma. 
  (Note that annihilation and Compton scattering 
  processes are associated to mean free paths $\propto E_\ini\,$, 
  and are thus suppressed 
  when $E_\ini \to \infty$ for a finite-size QGP.) 
} 
In a single scattering with the medium, 
the heavy quark initial and final four-momenta will be denoted by 
$\P_\ini = (E_\ini\,, \vec{p}_\ini)$ 
and 
$\P_\fin = (E_\fin\,, \vec{p}_\fin)$. 
The four-momentum transferred to the medium is then 
$\P_\ini - \P_\fin \, \equiv \, {\cal Q} = (\eps, \vec{q})$, 
of virtuality $\Q^2 = \eps^2 - \vec{q}^2 = t\,$, 
with $t$ the Mandelstam variable in the elementary elastic scattering. 
We will consider the $E_\ini \to \infty$ limit, 
other scales being finite (in particular $\eps$ and $|\vec{q}|$), 
which implies $|t| \ll |t|_{\rm max} \to \infty$.\footnote{%
  Using the Mandelstam variables defined in \eq\nr{eq:stu}, 
  we have $|t|^{ }_\rmi{max} = (s-M^2)^2/s$. 
  For a given momentum $k$ of the thermal parton participating 
  in the elastic scattering,  
  the maximum $|t|$ is obtained 
  by setting $s \to M^2+ 4 E_\ini\,k$ 
  (corresponding to a head-on collision in the plasma rest frame). 
  }
It follows that the elastic scattering of the heavy quark 
in the medium is dominated by $t$-channel exchange. 
The energy transfer then coincides 
with the longitudinal momentum transfer,
\be
  \eps 
  \; = \; 
  E^{ }_\ini - E^{ }_\fin  
  \; = \; 
  \sqrt{ \vec{p}^2_\ini +M^2 } 
  - \sqrt{ (\vec{p}^{ }_\ini - \vec{q})^2 +M^2 } 
  \; \simeq \;
  v \, q^{ }_{\parallel} \ +\, \rmO\big(1/E_\ini\big) 
  \; ,  
  \la{q0-approx}
\ee
where $v \equiv p_\ini / E_\ini \to 1$ is the heavy quark velocity, 
and the longitudinal and transverse components of $\vec{q}$ are defined 
w.r.t. the initial heavy quark direction, 
$
  \vec{q} 
  \;=\; 
  \vec{q}^{ }_\perp + q^{ }_{\parallel} \, 
  \vec{p}^{ }_\ini \big/ p^{ }_\ini \,
$. 

Let us recall that 
the approximation of small momentum transfer $|t| \ll s$ is sufficient 
to derive the average energy loss ${\rm d}E_\rmi{coll}/{\rm d}x$ 
in a leading-logarithmic approximation~\cite{Bjorken:1982tu,Thoma:1990fm}. 
Indeed, ${\rm d}E_\rmi{coll}/{\rm d}x$ depends on 
the first moment of $w(\eps)$, which to leading-log can be obtained 
by choosing an upper bound for $\eps$ that is proportional 
(though significantly smaller) to the maximum energy transfer 
$\eps_{\rm max}$ kinematically allowed in an elastic collision. 
In our study, the focus is not on the average loss, but 
on the distribution $f(x,\Delta)$, which can be derived in 
the strict $E_\ini \to \infty$ limit, implying $\eps_{\rm max} \to \infty\,$. 
Of course, the distribution 
will then have an undefined (infinite) average, 
but should be valid, 
in a practical situation where $E_\ini$ is finite, 
as long as $\Delta \ll E_\ini\,$. 
We will come back to this point in the final discussion, 
see sec.~\ref{se:disc}. 

The rate $w(\eps)$ in the small-$t$ approximation can be obtained by 
integrating over $\vec{q}^{ }_\perp$ a more general quantity, 
namely the fully differential elastic scattering rate, which reads
\bea
  \frac{ 
    {\rm d}\Rate
  }{
    {\rm d} \eps \, {\rm d}^2 \vec{q}^{ }_\perp
  }
  & \simeq &  {\frac{g^2 \CF}{(2\pi)^2} }
  \, \left[ 1 + \nB\big(\eps\big) \right] \, 
  \frac{q_\perp^2}{\eps^2+q_\perp^2} \, 
  \big[\, \rho^{ }_\rmii{T} (\eps, q) - \rho^{ }_\rmii{L} (\eps, q) \,\big]
  \; , 
  \la{eq:gamma_el}
\eea
where $g^2 = 4 \pi \alpha_s$ is the gauge coupling, 
$\CF = (\Nc^2-1)/(2\Nc)$ the heavy quark Casimir charge, 
$\nB$ the Bose-Einstein distribution, 
and $\rho_\rmii{T,L}^{ }$ are the transverse and longitudinal 
gluon spectral densities, which by virtue of \eq\nr{q0-approx} 
should be evaluated at $q=(q_\perp^2 + \eps^2)^{1/2}\,$. 
The expression \nr{eq:gamma_el} 
is obtained by setting $v \to 1$ in \eq\nr{eq:gamma_el_exact}, 
derived in appendix~\ref{app:xsection}. 

Let us note that the differential rate 
${{\rm d} \Rate}/{{\rm d}^2 \vec{q}^{ }_\perp}$ 
(the $\eps$-integrated version of \eq\nr{eq:gamma_el}), 
which has been determined to leading-order~\cite{Arnold:2008vd} 
and beyond~\cite{Caron-Huot:2008zna}, 
is of primary interest in many studies of jet-quenching, 
radiative energy loss and $p^{ }_\perp$-broadening. 
In those contexts, transverse momentum diffusion can be described by a 
kinetic equation~\cite{Blaizot:2013vha,Boguslavski:2024jwr}
of similar structure to the one governing 
the collisional quenching weight discussed here, involving the 
transport coefficient $\hat q$ 
characterising transverse momentum broadening in the target, defined as
\be
  \la{eq:qhat}
  \hat q 
  \; = \; 
  \int {\rm d}^2 \vec{q}^{ }_\perp \, 
  \frac{ {\rm d} \Rate }{{\rm d}^2 \vec{q}_\perp}
  \; q_\perp^2  \ . 
\ee
By contrast, our study needs the 
$\vec{q}^{ }_\perp$-integrated differential scattering rate, 
which is less commonly discussed. 
Integrating \nr{eq:gamma_el} over $\vec{q}^{ }_\perp$ 
and changing variable from $q_\perp^2$ to $q^2=q_\perp^2+\eps^2$, 
the differential rate for energy transfer $\eps$ 
takes the general form 
\be
  \la{eq:w-eps-general}
  w(\eps) 
  \ \equiv \ \frac{{\rm d}\Rate}{{\rm d} \eps}
  \; = \;
  \alpha_s \CF \,  [ 1 + \nB\big(\eps \big) ] \; \Sfunc(\eps)  
  \, , 
  \ee
  where
  \be
  \la{eq:Sofeps}
  \Sfunc\big(\eps)
  \; = \; 
  2
  \int_{|\eps|}^{\infty} \dd q  
  \ q \, 
  \left(
   1-  \frac{\eps^2}{q^2} 
    \right)
  \big\{ \rho_\rmii{T}(\eps,q) -  \rho_\rmii{L}(\eps,q) \big\} 
  \; \equiv \;
  \Sfunc^{ }_\rmii{T}(\eps) 
  + 
  \Sfunc^{ }_\rmii{L}(\eps)  
  \ . 
\ee
The functions $\Sfunc_i(\eps)$ 
(for $i=\{\rmi{T},\rmi{L}\}$) are odd functions of $\eps\,$, 
following from the property 
$\rho_i^{ }(-\eps,q) = - \rho_i^{ }(\eps,q)$ of gluon spectral densities 
(see appendix~\ref{app:xsection}).
Note that eqs.~\nr{eq:w-eps-general}--\nr{eq:Sofeps} 
are quite general, and could apply to various physical situations, 
associated with different spectral gluon densities.

%
\subsection{Solution for the energy loss distribution}
\label{sec:kin-eq}

In this section we first recall in \eq\nr{eq:bromwich} 
the general solution to the kinetic equation~\nr{eq:kinetic1}, 
and then discuss some of its features for the rate $w(\eps)$ 
of the form \nr{eq:w-eps-general}. 
Here we only assume that 
the odd function $\Sfunc(\eps)$ is integrable at $\eps \to \infty\,$.
(This will be the case for the perturbative QCD medium 
considered in sec.~\ref{sec:perturbative-rate}, 
where $\Sfunc(\eps) \sim 1/\eps^2$ at large $\eps\,$, 
see \eq\nr{eq:S_hard_expansions}.)

Introducing the Laplace transform of $f(x,\Delta)$ (w.r.t. the 
variable $\Delta$), and then making use of the inverse Laplace transform, 
the solution to \eq\nr{eq:kinetic1} with initial condition 
$f(0,\Delta) = \delta(\Delta)$ is given formally by\footnote{%
  Equation~\nr{eq:bromwich} is identical to eq.(5) of
  ref.~\cite{Landau:1944fvs}, 
  with the exception that the integral over $\eps$ also receives 
  a contribution from $-\infty$ to $0$ in our case, 
  where energy gain is possible. 
  In the following, even in the presence of thermal effects (when $T>0$) 
  leading to energy gain $\Delta < 0\,$, 
  for simplicity we will refer to $f(x,\Delta)$ as 
  the ``energy loss'' distribution.
} 
\be
  f(x,\Delta) 
  \; =\; 
  \frac{1}{2\pi i } 
  \int_{\cal B} \, \dd \nu \;
  \exp 
  \biggl\{\,
  \nu \Delta - x 
 \underbrace{ \int_{-\infty}^\infty {\rm d}\eps\,
  w(\eps) \big( 1- e^{-\nu\eps} \big) 
  }_{ \displaystyle \equiv I(\nu) } \,
  \, \biggr\} \ 
\la{eq:bromwich} 
\ee
where the integration contour ${\cal B}$ 
in the complex $\nu$-plane, commonly 
called a {\it Bromwich contour}, 
is parallel to the imaginary axis and intersects 
the real axis at a certain point $\nu_0\,$. 
Inserting \nr{eq:w-eps-general} in \nr{eq:bromwich}, 
we see that the function $I(\nu)$ (and thus $f(x,\Delta)$) 
is well-defined provided $\nu_0$ is chosen as 
$0 \leq \nu_0 \leq \beta \equiv \frac{1}{T}$, 
ensuring convergence of the $\eps$-integral at $\pm \infty$.

To fuel the discussion below, let's introduce the heavy quark 
{\it damping rate},\footnote{%
  Historically, calculations of $\Gamma$ played an important 
  part in the development of HTL technology 
  in thermal field theory.  
  In hot gauge theories, the fermion damping rate is 
  (logarithmically) 
  infrared divergent 
  due to unscreened magnetostatic 
  gauge fields~\cite{Braaten:1992gd,Pisarski:1993rf}. 
  However, for cold dense systems, the absence of Bose-Einstein 
  enhancement (for gluon exchanges with $\eps \simeq 0$)
  renders the damping rate finite~\cite{Vanderheyden:1996bw,LeBellac:1996kr}.
}
  \be
  \la{damping-rate}  
  \Rate  
  \; = \;  
  \int_{-\infty}^{\infty} 
  {\rm d} \eps \, w(\eps) 
  \; = \; 
  \mathop{\rm lim}_{\nu \to \pm i \infty} I(\nu) \ . 
\ee
Depending on the precise form of $\Sfunc(\eps)$ used in the interaction rate 
$w(\eps)$ (see \eq\nr{eq:w-eps-general}), 
the damping rate might have some (infrared) divergence at $\eps \to 0$. 

When there is no such infrared divergence and $\Rate$ is finite, we can write  
\be
  e^{- x I(\nu)} 
  \ = \ 
  e^{- x  \Rate} \  e^{x \int {\rm d} \eps\, w(\eps) \,e^{-\nu\eps} } 
\la{expexp}
\ee
and expand the second exponential factor as a series, 
to rewrite \eq\nr{eq:bromwich} formally as 
\be
\la{f-Poisson}
  f(x,\Delta)
  \; = \; 
  e^{- x  \Rate} \,
  \sum_{n=0}^\infty \ 
  \frac{x^n}{n!} 
    \left[\,
          \prod_{i=1}^n
          \int_{-\infty}^{\infty}  \dd \eps^{ }_i \, w(\eps^{ }_i)
  \,\right] 
  \, 
  \delta
  \bigg(
          \Delta - \sum_{i=1}^n \eps^{ }_i 
  \bigg) 
  \; . 
\ee
Although the representation~\nr{f-Poisson} is less compact than 
the formal solution in \eq\nr{eq:bromwich}, it offers 
a more transparent physical interpretation. 
It shows that the energy transfer 
mechanism under consideration is a Poisson process, whereby successive 
transfers are statistically independent. 
The term with $n$ elastic collisions contributes 
to $f(x,\Delta)$ with a weight $e^{- x \Rate} \, \frac{(x \Rate)^n}{n!}\,$, 
which is the probability that exactly 
$n$ scatterings occur over the distance $x\,$. 
In particular, the $n=0$ term corresponds to the absence of scattering
(occurring with probability $e^{ - x \Rate}$), 
and yields a contribution $e^{ -x \Rate} \delta (\Delta)$ 
in $f(x,\Delta)$. 

This ``no-scattering'' component of the distribution can also be identified 
directly from the integral representation~\nr{eq:bromwich}. 
There, it arises from the 
limit $\nu \rightarrow \pm i \, \infty\,$, 
where the integrand behaves as $e^{ \nu \Delta } e^{ - x \Rate}$. 
%
%
The Bromwich integral must therefore 
be understood in the sense of a distribution 
when $\Gamma$ is finite, and the singular (Dirac-$\delta$) 
part can be isolated as 
\be
  \la{f-delta-piece}
  f(x,\Delta) 
  \; = \; 
  e^{- x \Rate} \, \delta(\Delta) 
  \, + \, 
  \int_{\cal B} \, 
  \frac{{\rm d} \nu}{2\pi i } \ 
  \left[ \,     e^{ \nu \Delta - x  I(\nu) }
              - e^{ \nu \Delta - x \Rate   } \, \right] \ ,
\ee
where the remaining integral is convergent 
and represents the contribution from all 
configurations involving {\em at least} one scattering.

The situation changes drastically if $\Rate$ diverges, 
e.g.\ due to the infrared behaviour of $w(\eps)$. 
In this case, the probability $e^{-x \Rate}$ of having no scattering vanishes 
for any finite $x\,$, as does the probability of 
undergoing any {\em finite} number of energy transfers.\footnote{%
  The test particle experiences 
  infinitely many scatterings transferring an infinitesimal energy, 
  such that the 
  total energy exchange $\Delta$ remains finite. 
  This situation is similar to multiple soft photon emission 
  contributing to the finite energy loss of an energetic 
  charge~\cite{Dokshitzer:1995}.
} 
As a consequence, in this regime there is no $\delta(\Delta)$ term 
in $f(x,\Delta)$ (except obviously at $x =0$ where 
$f(0,\Delta) = \delta(\Delta)$). 
Despite that \eq\nr{f-Poisson} becomes formally ill-defined, 
and thus impractical when $\Rate = \infty\,$, it still 
reflects the underlying Poissonian structure. 
The infinite series in \eq\nr{f-Poisson} should in this case be 
regarded as a formal expansion of the general 
solution~\nr{eq:bromwich}. 
The latter 
may remain well-defined 
even for $\Rate = \infty\,$, 
provided the factor $(1-e^{-\nu \eps})$ appearing in $I(\nu)$ 
is sufficient to tame the infrared behaviour of $w(\eps)\,$. 

It will prove convenient to rewrite 
the function $I(\nu)$ defined by \eq\nr{eq:bromwich} 
by changing variable $\eps \to -\eps$ in 
the part of the integral with $\eps <0\,$, 
\ie\ corresponding to energy gain. 
Using $1+\nB(-\eps) = -\nB(\eps)$, 
and recalling that $\Sfunc(-\eps)=-\Sfunc(\eps)$, we obtain 
\bea
\la{eq:Inu-2}
  I(\nu) 
  \, = \, 
  \alpha_s \CF \int_{0}^\infty 
  \dd\eps\, \Sfunc(\eps) \, 
  \Bigl\{ 
  [ 1 + \nB\big(\eps \big) ]  \big( 1- e^{-\nu\eps} \big) 
  + 
  \nB\big(\eps \big)   \big(1- e^{\nu\eps} \big) 
  \Bigr\} \hskip 18mm  && 
  \\[1mm]
  \la{eq:Inu-3}
   \, = \, 
  \alpha_s \CF \left[ \, 
  \int_{0}^\infty 
  \dd\eps\, \Sfunc(\eps) \, \big( 1- e^{-\nu\eps} \big) 
  +  2 \int_{0}^\infty \dd\eps\, \Sfunc(\eps) \, \nB\big(\eps \big)
  \big(1- \cosh{\nu\eps} \big) 
  \, \right] \,,  \hskip 4mm && 
\eea
where the first term in the latter bracket determines 
the $T \to 0$ limit of $I(\nu)$.\footnote{%
  \la{foot:terminology}In general, the first term of \eq\nr{eq:Inu-3} 
  may depend on the temperature, 
  but should remain finite when $T \to 0$ 
  (unlike the second term, which vanishes in this limit). 
  This is what will happen in the specific case examined 
  in sec.~\ref{sec:perturbative-rate}, 
  where $\Sfunc(\eps)$ depends on $T$ through the Debye mass 
  $\mD^{ }$ defined by \eq\nr{eq:mD}. 
  By considering non-zero chemical potential, we can take $T=0$ 
  with finite $\mD^{ }\,$.
  }
The collisional quenching weight $f(x,\Delta)$ 
is thus obtained by inserting \eq\nr{eq:Inu-3} 
in \eq\nr{eq:bromwich}, 
and performing the $\nu$-integral along the Bromwich contour ${\cal B}$. 
%

%
\section{Differential scattering rate in perturbation theory}
\label{sec:perturbative-rate}

In this section, we explicitly derive the function $\Sfunc(\eps)$ 
that appears in \eq\nr{eq:w-eps-general}, 
in the perturbative framework of field theory at 
finite temperature $T$ and/or density $\mu\,$. 
For this discussion it is convenient to introduce an intermediate scale 
$\eps^\star$ separating the ``soft scale'' 
$\mD$ (the Debye mass, see below) and the ``hard scale'' 
$\Lambda \equiv {\rm max}(\pi T, \mu)$, 
namely, $\mD \ll \eps^\star \ll \Lambda$, 
but being otherwise arbitrary~\cite{Braaten:1991dd}.
When addressing energies in the domain $\eps \sim \mD$, 
standard perturbation theory breaks down and must be resummed. 
A systematic approach involves the resummation of 
hard thermal loops (HTL)~\cite{Pisarski:1988vd,Frenkel:1989br,Braaten:1989mz}. 
In sec.~\ref{sec:S-soft} we derive $\Sfunc(\eps)$ for 
$|\eps| \ll \eps^\star$ 
using the explicit forms of $\rho^{ }_\rmii{T}$ and $\rho^{ }_\rmii{L}$ 
obtained within the HTL framework.\footnote{%
  For systems at finite density, the HTL 
  approximation is sometimes called 
  the hard dense loop (HDL) approximation. 
  For simplicity we shall use HTL for both thermal and dense systems.
} 
We will first recall the necessary ingredients 
for screening of the soft $t$-channel exchange, 
and then explain how the $q$-integral in \nr{eq:Sofeps}, 
when expressed in terms of HTL gluon spectral densities, 
can be replaced using Cauchy's theorem 
by a sum over poles in the {\em timelike} domain 
(the poles being solutions of the well-known thermal dispersion relations). 
This leads to the compact expression \nr{S-eps-HTL} which 
is easy to evaluate numerically. 
In sec.~\ref{sec:S-hard} we calculate $\Sfunc(\eps)$ for 
hard transfers $|\eps| \geq \eps^\star$ 
within standard kinetic theory, 
which yields the result in closed form \nr{S-eps-hard}. 
The matching of the expressions for $\Sfunc(\eps)$ obtained in 
the soft and hard regions is discussed in sec.~\ref{sec:matching}. 
Readers who are not interested in the technical details 
regarding the construction of $\Sfunc(\eps)$ might skip directly to 
sec.~\ref{sec:quenching-weight}~(p.~\pageref{sec:quenching-weight}).

%
\subsection{Main steps of computation in HTL domain}
\label{sec:S-soft}

Starting from a covariant and real-time formalism, 
the transverse and longitudinal gluon propagators can be 
expressed quite generally in terms of gluon self-energies 
$\Pi_\rmii{T,L}^{ }(\Q)$ as~\cite{Weldon:1982aq}
\be
 \Delta_\rmii{T,L}^{ }(\Q) 
 \; = \; 
 \frac{1}{-\Q^2 + \Pi_\rmii{T,L}^{ }(\Q)} 
 \; = \; 
 \frac{1}{ -\eps^2  + q^2 + \Pi_\rmii{T,L}^{ }(\eps, q)} 
 \  . 
\la{propagators-0}
\ee
The associated gluon spectral densities, entering \eq\nr{eq:Sofeps}, 
are defined by 
\be
  \rho_\rmii{T,L}^{ }(\Q) 
  \; = \; 
  \frac{1}{\pi} \im \Delta_\rmii{T,L}^{ }(\eps+ i 0^+, q)  \ ,
 \la{cut-0}
\ee
where the $i 0^+$ prescription specifies the analytic continuation 
from Euclidean to Minkowski frequencies 
(cf. appendix~\ref{app:xsection}).\footnote{%
  This prescription also determines the signs of the 
  imaginary parts of
  the electric permittivity and magnetic permeability 
  of the medium in accordance with general analytic properties, 
  see \eg\ \S77-82 of ref.~\cite{LL8}.
} 

In gauge theories at finite temperature or density, 
a $t$-channel soft gluon exchange (\ie, with $|\eps| \leq \eps^\star$) 
leads to infrared 
divergences in the naive weak-coupling expansion. 
These are regulated by HTL 
resummation~\cite{Braaten:1989mz,Taylor:1990ia,Blaizot:2001nr} 
which systematically includes corrections to 
the gauge boson propagator arising from physical effects 
like Debye screening and Landau damping.\footnote{%
  HTL corrections to the vertices~\cite{Braaten:1991gm} do not need 
  to be included here, since all external particles participating in 
  the elastic scattering have ``hard'' momenta $\gsim \max(\pi T, \mu)$. 
} 
The transverse and longitudinal HTL self-energies read
\bea
  \Pi^{ }_\rmii{T} (\Q) 
  &=&
  \frac{ \mD^2 }{2}
  \biggl\{\frac{\eps^2 }{q^2} + \frac{\eps}{2 q} 
  \biggl( 1 - \frac{\eps^2}{q^2} \biggr)
    \log \frac{\eps+q }{\eps - q } \biggr\} \; ,
  \la{eq:HTL_T}
  \\[2mm]
  \Pi^{ }_\rmii{L} (\Q) 
  &=&
  \mD^2  \biggl( 1  - \frac{\eps^2}{q^2} \biggr)
  \biggl\{1-\frac{\eps}{2 q} \log \frac{\eps+q }{\eps - q } \biggr\} \; ,
  \la{eq:HTL_L}
\eea
where the Debye mass 
is given by the thermal mass of 
colour-electric fields~\cite{Kapusta:1979fh,Laine:2016hma}
\be
\la{eq:mD}
  \mD^2 
  \; = \; 
  g^2  \mu^2 \frac{\nf}{2 \pi^2}
  \, + \,
  g^2  T^2 \, 
  \biggl( \frac{\Nc}3 + \frac{\nf}6 \biggr) 
  \, + \, 
  \rmO ( g^3 T^2 ) 
  \; .
\ee
(The omitted corrections are sensitive to the 
magnetic sector of QCD~\cite{Braaten:1994pk}.)
We thus view $T$ and $\mu$ as 
independent parameters which together 
fix $\mD$, allowing us to explore 
different screening regimes. 
For instance, we can consider the $T \to 0$ limit in \eq\nr{eq:Inu-3}, 
while keeping $\mD$ finite 
(as anticipated in footnote~\ref{foot:terminology}). 

The HTL spectral densities obtained 
from \eqs\nr{propagators-0} and \nr{cut-0}, 
using the self-energies in \eqs\nr{eq:HTL_T} and \nr{eq:HTL_L}, 
provide an explicit expression of $\Sfunc(\eps)$ 
(defined by \nr{eq:Sofeps}) 
valid in the soft domain $|\eps| \leq \eps^\star$. 
This expression for $\Sfunc(\eps)$ probes 
the spectral functions in the spacelike domain, 
but can be conveniently reformulated in terms of 
the poles of the gluon propagators in the timelike domain as follows. 

First, noting that the HTL self-energies are functions 
of the ratio ${\eps}/{q}$ only, 
in the spacelike domain $|\eps| <q$ 
the imaginary part in \nr{cut-0} can be written as 
(with $i=\{\rmi{T},\rmi{L}\}$) 
\be
  \im \Delta^{ }_i(\eps + i 0^+, q) 
  \; = \;  
  \im \Delta^{ }_i(\eps, q - i 0^+) 
  \; = \;
  -\frac{1}{2 i} {\rm \,Disc\,} \Delta^{ }_i(\eps, q) \ , 
\ee
where 
$
  {\rm Disc\,} \Delta_i(\eps, q) 
  \equiv 
  \Delta_i(\eps,q+ i 0^+)  - \Delta_i(\eps,q - i 0^+)
$ 
is the discontinuity of $\Delta_i^{ }(\eps, q)$, 
viewed as a function of the complex variable $q$, 
across the real $q$-axis.
Thus, at fixed $\eps >0$ the transverse contribution to 
\eq\nr{eq:Sofeps} can be written as
\be
 \la{ST-contourC}
  \Sfunc^{ }_\rmii{T}(\eps) 
  \, = \, 
  - \frac{2}{2 \pi i}
  \int_{\eps}^{\infty} \dd q  \ q \, 
  \left(
  1-  \frac{\eps^2}{q^2} 
  \right) 
  \, {\rm Disc\,} \Delta^{ }_\rmii{T}(\eps, q) 
  \, = \,  
  - \frac{1}{2 \pi i} \int_{\cal C}  \dd q  \ q \, 
  \left(
  1-  \frac{\eps^2}{q^2} 
  \right) \Delta^{ }_\rmii{T}(\eps, q) \ , 
\ee
where we used the fact that the function ${\rm Disc\,} \Delta^{ }_\rmii{T}(q)$ 
is odd in $q\,$, 
allowing one to replace 
$2  \int_{\eps}^{\infty} \to  \int_{-\infty}^{-\eps} +  \int_{\eps}^{\infty}$, 
and where $\cal C$ is the contour 
of the complex $q$-plane surrounding the 
semi-infinite intervals $\left( -\infty, -\eps \right]$ 
and $\left[ \eps, \infty \right)$ of the real axis. 
By closing the contour $\cal C$ with semicircles 
(in the upper and lower half-planes), 
and using 
$\Delta_\rmii{T}^{ }(\eps, q \to \infty) \simeq 1/q^2$, 
we obtain 
\be
  \la{ST-residues}
  \Sfunc^{ }_\rmii{T}(\eps) 
  \; = \;  
  1 - 
  \sum_\rmi{poles} {\rm Res\,} 
  \left\{ q \, 
  \left(1-  \frac{\eps^2}{q^2} \right) 
  \Delta_\rmii{T}^{ }(\eps, q) \right\}  \ ,
\ee
where the first term arises from the $q$-integral on the semi-circles. 
In the same way, for the longitudinal contribution to $\Sfunc(\eps)$ we get 
(note that $\rho^{ }_\rmii{L}$ comes with a minus sign in \eq\nr{eq:Sofeps}) 
\be
  \la{SL-residues}
  \Sfunc^{ }_\rmii{L}(\eps) 
  \; = \;  
  - 1 + 
  \sum_\rmi{poles} {\rm Res\,} 
  \left\{ q \, 
  \left(1-  \frac{\eps^2}{q^2} \right) 
  \Delta_\rmii{L}^{ }(\eps, q) \right\}  \ . 
\ee

In each of eqs.~\nr{ST-residues} and \nr{SL-residues}, 
the sum of residues is taken over three poles 
of the relevant function, 
namely, one pole at $q=0$ and two poles of the gluon propagator 
$\Delta^{ }_i(\eps, q)$ 
located at $q = \pm q^{ }_i(\eps) \equiv \pm q^{ }_{i,\epsilon}$.\footnote{%
  In other words, $\pm q^{ }_{i,\epsilon}$ are 
  the solutions to the dispersion relation 
  $-\eps^2  + q^2 + \Pi^{ }_i(\eps, q) = 0$ for the variable $q\,$.
} 
The latter two poles lie on the real axis when 
$\eps > \frac{ \mD }{ \sqrt{3}}\,$, 
in which case they correspond to the 
(transverse or longitudinal) plasmon modes, 
and on the imaginary axis when 
$\eps < \frac{ \mD }{ \sqrt{3} }$~\cite{Weldon:1982aq}. 
By evaluating the residues at the various poles we obtain: 
\bea
\la{ST-eps-HTL}
  \Sfunc^{ }_\rmii{T}(\epsilon) 
  & = &  
  1 
  - \frac{3 \eps^2}{3 \eps^2-\mD^2} 
  - \frac{
    2\big( \epsilon_{ }^2 - q^2_{\rmii{T},\epsilon} \big)^2
  }{
    3\big( \epsilon_{ }^2 - q^2_{\rmii{T},\epsilon} \big)^2 
    - \epsilon_{ }^2 \mD^2
  }  \ , \\[2mm]
 \la{SL-eps-HTL}
 \Sfunc^{ }_\rmii{L}(\epsilon) 
 & = &  
 - 1 
 + \frac{3 \eps^2}{3 \eps^2-\mD^2} 
 + \frac{
   2\big(\epsilon_{ }^2 - q^2_{\rmii{L},\epsilon}\big)
 }{
   3\big(\epsilon_{ }^2 - q^2_{\rmii{L},\epsilon}\big) - \mD^2
 } \ , 
\eea
where in each expression the second term arises from the pole at $q=0$, 
and the third term from the two poles at $q = \pm q^{ }_{i,\epsilon}$ 
(which have the same residue).\footnote{%
  The residues of the poles at $q=\pm q^{ }_{i,\epsilon}$ 
  are obtained as follows. 
  Close to the pole at 
  $q = +q^{ }_{i,\epsilon}\,$, the function 
  $
    \Delta_i^{-1}(\eps, q) 
    = 
    -\eps^2  + q^2 + \Pi_{i}^{ }(\eps,q) 
    = 
    - \eps^2 (1-y^2) + \Pi_{i}^{ }(\eps, \eps\,y) 
    \equiv 
    g^{ }_{i}(y)
  $ 
  (with $y \equiv \frac{q}{\eps}$), 
  can be expanded as $g^{ }_i(y) \simeq (y - y^{ }_i) g_{i}^\prime(y^{ }_i)$, 
  where $y_i \equiv \frac{q_{i,\epsilon}}{\eps}$ and using $g^{ }_{i}(y_i) = 0$. 
  Noting that $g^{ }_{i}$ satisfies the differential equation 
  $
    y(y^2-1) \, g_{i}^\prime(y)
    =
    (3-y^2) \, g^{ }_{i}(y) + 3(y^2-1)^2 \eps^2-1
  $ 
  for 
  $i = \rmi{T}$ and 
  $
    y (1-y^2) \, g_{i}^\prime(y)
    =
    (y^2-1) \, g^{ }_{i}(y)+y^2 [ 3(1-y^2) \eps^2-1 ]
  $ 
  for $i = \rmi{L}\,$, 
  the factor $g_{i}^\prime(y^{ }_i)$ 
  is directly obtained by setting 
  $y \to y^{ }_i$ in those equations (and using $g^{ }_{i}(y^{ }_i) = 0$).
}
Let us remark that when combining 
$\Sfunc^{ }_\rmii{T}$ and $\Sfunc^{ }_\rmii{L}\,$, 
the contributions from the pole at $q=0$ cancel out, 
and we obtain the quite compact expression: 
\be
\la{S-eps-HTL}
  \Sfunc^{ }_\rmii{HTL}(\epsilon) 
  \; = \; 
  \frac23 
  \, \bigg[\,
    \frac{ \mD^2 }{
      3\big(\epsilon_{ }^2 - q^2_{\rmii{L},\epsilon}\big)
      - \mD^2
    }
    \; - \;
    \frac{ (\epsilon\, \mD)^2 }{
      3\big(\epsilon_{ }^2 - q^2_{\rmii{T},\epsilon}\big)^2
      - (\epsilon\, \mD )^2
    }
  \, \bigg] \ , 
\ee
where we recall that $\pm q^{ }_{i,\epsilon}$ are the solutions 
to the dispersion relation $\epsilon^2 = q^2 + \Pi^{ }_i (\epsilon, q)\,$. 
This result is analogous to a famous ``sum rule'' derived 
for the differential rate in transverse momentum 
${\rm d}\Rate/{\rm d}^2 \vec{q}^{ }_\perp$, obtained 
by integrating the fully differential rate 
\nr{eq:gamma_el} over $\eps$~\cite{Aurenche:2002pd}. 

%
\begin{figure}[t]
\centerline{
    \includegraphics[scale=.65]{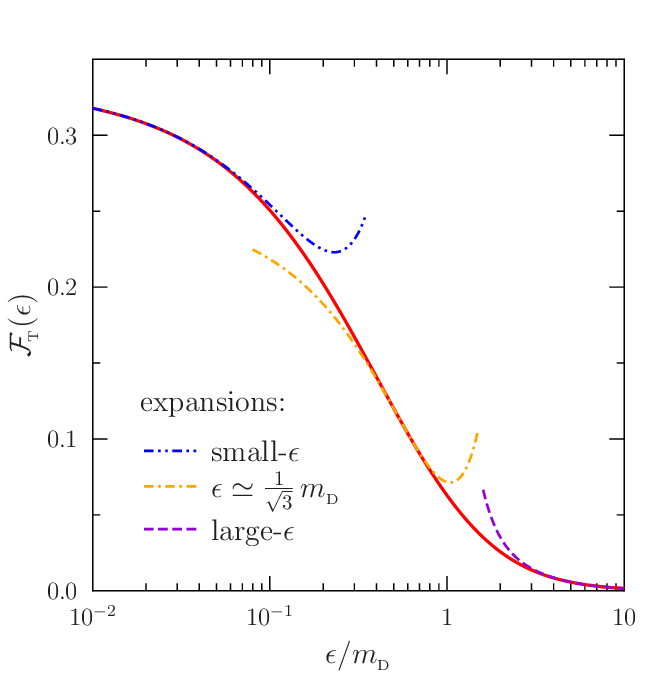}
  ~~\includegraphics[scale=.65]{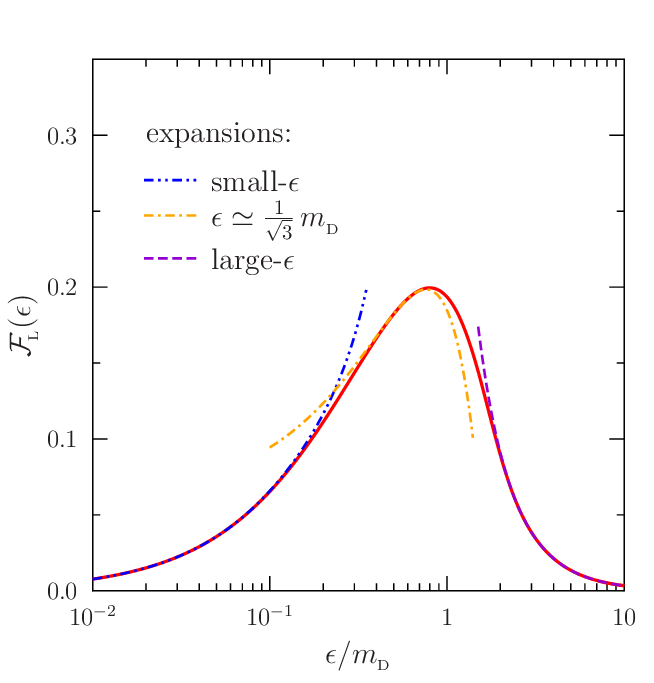}
}
\vspace*{-1mm}
\caption[a]{\small 
  Functions $\Sfunc^{ }_\rmii{T}(\eps)$ (left) and 
  $\Sfunc^{ }_\rmii{L}(\eps)$ (right) 
  defined by \eq\nr{eq:Sofeps} and using HTL gluon spectral densities. 
  These functions are compared locally to their respective 
  asymptotic expansions at 
  $\eps \ll m_{\rmii{D}}\,$, $\eps \simeq \frac{m_{\rmii{D}}}{\sqrt{3}}\,$, 
  and $\eps \gg m_{\rmii{D}}$ up to the 
  order given in appendix~\ref{app:Taylor-ex}.
}
\la{fig:S}
\end{figure}
%

The expressions in eqs.~\nr{ST-eps-HTL} and \nr{SL-eps-HTL} 
have the following nice features: 
(i) Since only $q^2$ appears in the formulae, 
these expressions naturally include the case when $q$ is purely imaginary. 
(ii) They prove better than 
\eq\nr{eq:Sofeps} for an efficient numerical evaluation, 
since they do not require an integral over $q\,$, 
but just a {\em local} solution to an implicit equation. 
Numerically, we can obtain $\Sfunc^{ }_\rmii{T,L}(\epsilon)$ 
to a desired accuracy by finding the 
timelike roots to that accuracy.
(iii) Using eqs.~\nr{ST-eps-HTL} and \nr{SL-eps-HTL}, 
one can obtain asymptotic expansions of $\Sfunc^{ }_\rmii{T,L}(\eps)$ 
(to any order) at 
$\eps \ll \mD$, 
$\eps \simeq \frac{ \mD }{ \sqrt{3} }$, 
and 
$\eps \gg \mD\,$. 
These expansions are given in appendix~\ref{app:Taylor-ex}.
The functions $\Sfunc^{ }_\rmii{T}(\eps)$ and $\Sfunc^{ }_\rmii{L}(\eps)$, 
together with their asymptotic expansions, are shown in fig.~\ref{fig:S}. 

Note that $\Sfunc_\rmii{T}(\eps)$ remains finite when $\eps \ll \mD\,$, 
$\Sfunc_\rmii{T}(\eps) \simeq \frac{1}{3}$ (see~\eq\nr{ST_expansions}). 
This leads to an apparent discontinuity of $\Sfunc_\rmii{T}$ at $\eps=0$ 
(since $\Sfunc_\rmii{T}(\eps)$ is necessarily an odd function), 
and to the infrared divergence of the total rate \nr{damping-rate} 
as soon as $T>0$ (due to $\nB\big(\eps \big) \simeq T/\eps$ when $\eps \to 0$).
However, $\Sfunc^{ }_\rmii{T,L}(\eps)$ have been obtained within 
the HTL framework, and should thus only be used in 
the domain where $\eps \sim \mD \sim g \Lambda\,$. 
The apparent discontinuity of $\Sfunc_\rmii{T}$ at $\eps=0$ 
should be cured beyond HTL, 
by non-perturbative magnetic screening effects appearing at 
the ``ultrasoft'' scale $g^2 T \ll \mD$~\cite{Linde:1980ts} 
(the so-called Linde problem),
whereby $\Sfunc_\rmii{T}(\eps \to 0) = 0$
and thus the total rate becomes finite. 

In our study, the quantity of interest is
not the total rate, but the probability distribution 
$f(x,\Delta)$ (cf.~\eq\nr{eq:bromwich}). 
The latter depends on the function 
$I(\nu)$ which has no infrared singularity 
(thanks to the factor $1- e^{-\nu\eps}$). 
For our concern, the divergence of the total rate and 
the spurious discontinuity of $\Sfunc_\rmii{T}(\eps)$ at $\eps =0$, 
occurring within the HTL framework, should be innocuous. 

Of course, the HTL approximation also fails when $\eps$ 
becomes similar to (or exceeds) the hard scale 
$\Lambda = {\rm max}(\pi T, \mu) \gg \mD\,$, 
due to oversimplified kinematics. 
In this ``hard domain'', the functions $\Sfunc_\rmii{T,L}$ describe Born level 
$2\to 2$ scattering of the fast heavy quark 
with a gluon or light quark 
in the medium. 
This is addressed in the next section.

%
\subsection{Scattering rate for hard energy transfer}
\label{sec:S-hard}

While \eq\nr{eq:Sofeps} provides a general representation which 
remains valid for hard energy transfers, 
in this domain it is more convenient to obtain 
the heavy quark interaction rate directly from kinetic theory, namely,
\bea
\left. \frac{{\rm d}\Rate}{{\rm d} \eps} 
  \right|_{\rm hard}
  \ = \ 
  \frac{1}{2\Nc} \cdot \frac{1}{2 E^{ }_\ini }
  \int
 {\rm d}\Omega^{ }_{2\to2} \ \delta\big(\eps + E^{ }_\fin -E^{ }_\ini \big) \  
 \bigg\{\,
   \nB\big( k^{ }_\ini \big)
   [1+\nB\big( k^{ }_\fin \big)] 
\,{\textstyle \sum} \, |{\cal M}_{g}|^2_{ } \hskip 8mm && 
  \nn[0.5mm]
   \, + \, 
   \nF\big( k^{ }_\ini - \mu \big)
   [1-\nF\big( k^{ }_\fin - \mu \big)] 
\, \nf \,{\textstyle \sum} \, |{\cal M}_{q}|^2_{ }  \hskip 8mm && 
  \nn[1mm]
   \, + \, 
   \nF\big( k^{ }_\ini + \mu \big)
   [1-\nF\big( k^{ }_\fin + \mu \big)] 
\, \nf \,{\textstyle \sum} \, |{\cal M}_{\bar q}|^2_{ }  \,\bigg\} \ ,  
  \hskip 2mm && 
  \la{hard-rate}
\eea
where $\nF$ is the Fermi-Dirac distribution, 
and the factor ${1}/(2\Nc)$ indicates that 
the rate is averaged over the initial heavy quark  spin and colour 
(consistently with \eq\nr{eq:gamma_el_0}). 

In this expression, $\sum |{\cal M}_{j}|^2_{ }$ denotes 
the squared matrix element (summed over spin and colour indices) 
for elastic scattering off a thermal parton $j=\{ g, q, {\bar q}\}$
(of initial and final momenta 
$\K^{ }_\ini = (k^{ }_\ini\,, \vec{k}^{ }_\ini\,)$ 
and 
$\K^{ }_\fin = (k^{ }_\fin\,, \vec{k}^{ }_\fin\,)$), 
$\nf$ is the number of light quark flavours in the plasma, 
and the two-body phase space integration measure is defined as
\be
 {\rm d}\Omega^{ }_{2\to2}  
 \; \equiv  \;
 \frac{
   {\rm d}^3\vec{p}^{ }_\fin 
 }{ (2\pi)^3 2 E^{ }_\fin } 
 \frac{ {\rm d}^3\vec{k}^{ }_\ini 
 }{ (2\pi)^3 2 k^{ }_\ini }
 \frac{ {\rm d}^3\vec{k}^{ }_\fin 
 }{ (2\pi)^3 2 k^{ }_\fin }
 \,(2\pi)^4 \delta^{(4)}\big( \P^{ }_\ini + \K^{ }_\ini
                            - \P^{ }_\fin - \K^{ }_\fin \,\big) 
  \; .
  \la{dO2to2} 
\ee
For the record, the Mandelstam variables of the elastic scattering are: 
\be
  \la{eq:stu}
  s \; \equiv \; (\P^{ }_\ini + \K^{ }_\ini)^2 \ \ ; \ \ \  
  t \; \equiv \; \Q^2  \; = \; (\P^{ }_\ini - \P^{ }_\fin)^2 \ \ ; \ \ \ 
  u \; \equiv \; (\P^{ }_\ini - \K^{ }_\fin)^2 
  \; . 
\ee

As mentioned in sec.~\ref{se:setup}, we work in the $v \to 1$ limit, 
and always assume $\eps \ll E_\ini\,$. 
In the latter limit of small energy transfer, 
%
%
which implies  $|t| \ll |t|^{ }_\rmi{max}\,$, 
the squared matrix element takes the same form for any type of target parton, 
\be
 {\textstyle \sum} \, |{\cal M}_{j}|^2_{ } 
 \; \overset{ }{\underset{_{ |t| \ll |t|^{ }_\rmiii{max}}}{\simeq}} \;
  c_j  \, 16 g^4 \CF \Nc \, \frac{(s-M^2)^2}{t^2} 
  \; , 
 \la{Mi-squared}
\ee
with 
$c_j =  \{ \Nc, \frac{1}{2}, \frac{1}{2} \}$ 
for 
$j = \{ g, q, {\bar q} \}\,$. 

When $v \to 1$ and $|t| \ll |t|_{\rm max}\,$, 
the calculation of 
$
  \left. \frac{{\rm d}\Rate}{{\rm d} \eps} \right|_{\rm hard}
$ 
can be performed exactly. Using \eq\nr{Mi-squared} 
together with the identities 
$
  \nB^{ }(x) [ 1 + \nB^{ }(x+y)] 
  = 
  [ 1 + \nB^{ }(y)][ \nB^{ }(x) - \nB^{ }(x+y)]
$
and
$
  \nF^{ }(x) [ 1 - \nF^{ }(x+y)] 
  = 
  [ 1 + \nB^{ }(y)][ \nF^{ }(x) - \nF^{ }(x+y)]
$, 
we can express 
$\left. \frac{{\rm d}\Rate}{{\rm d} \eps} \right|_{\rm hard}$ 
given by \eq\nr{hard-rate} in the form of \eq\nr{eq:w-eps-general}, 
and obtain the corresponding factor $\Sfunc^{ }_\rmi{hard} (\eps)$. 
Explicitly: 
\bea
  \la{Shard-0}
  \Sfunc^{ }_\rmi{hard} (\eps) 
  & = & 
  \frac{g^2}{(2\pi)^2} \, \int_{0}^\infty \! {\rm d} k^{ }_\ini  \, 
  k^{ }_\ini  \left[ \Phi(k^{ }_\ini ) - \Phi(k^{ }_\ini +\eps) \right] 
  \nn[0mm]
  & \times & 
  \int \frac{{\rm d} \Omega_{\vec{k}_\ini}}{4 \pi} 
  \int \frac{ {\rm d}^3\vec{k}^{}_\fin}{\pi}  \, 
  \frac{E_\ini }{E_\fin} \, 
  \delta \big( \eps - E_\ini + E_\fin  \big) \, 
  \frac{
    4 \big(       k^{ }_{\ini\phantom{\parallel}} \!\! 
                - k^{ }_{\ini\, \parallel} \big)^2
       }{t^2} \, 
  \frac{
          \delta(\eps + k^{ }_\ini  -k^{ }_\fin )
       }{2 k^{ }_\fin}
  \; ,  
  \hskip 5mm 
\eea
with $t = \eps^2 - (\vec{k}^{ }_\ini -\vec{k}^{ }_\fin)^2$ 
and the shorthand notation 
\be
\la{Phi-def}
  \Phi(k) 
  \; \equiv \; 
  2\Nc \, \nB(k) 
  + 
  \nf \, [ \nF(k - \mu)+\nF(k + \mu) ] 
  \; . 
\ee

In the second line of \eq\nr{Shard-0}, 
the integral over $k^{ }_{\fin\,\parallel}$ 
is performed using (see eq.~\nr{exact-delta})
\be
 \frac{E_\ini }{E_\fin} \, 
 \delta \big( \eps - E_\ini + E_\fin \big) 
 \; \simeq \; 
 \delta \big(\eps - q_\parallel \big) 
 \; = \;  
 \delta \big(\eps + k^{ }_{\ini\,\parallel} 
                  - k^{ }_{\fin\,\parallel} \big)  
 \; , 
\ee
so that the Mandelstam variable $t$ becomes 
$t = - (\vec{k}^{}_{\ini\,\perp} -\vec{k}^{}_{\fin\,\perp})^2$. 
The ${\rm d}^2 \vec{k}^{}_{\fin\,\perp}$ integral can be performed by 
integrating first over the azimuthal angle with $\vec{k}^{}_{\ini\,\perp}$, 
and then over $k^{}_{\fin\,\perp}$ using 
\be
 \frac{\delta(\eps + k^{ }_\ini  -k^{ }_\fin \,)}{2 k^{ }_\fin} 
 \; = \;
 \delta \Big( (\eps + k^{ }_\ini)^2 - k^{2}_\fin \Big)  
 \; = \;
 \delta \Big( 
    - k^{2}_{\fin\,\perp}+ k^{2}_{\ini\,\perp} 
    + 2 \eps \big( 
                      k^{ }_{\ini\phantom{\parallel}}\!\! 
                    - k^{ }_{\ini\,\parallel} 
             \big) 
 \Big)  \; .
\ee
After these elementary manipulations, 
the second line of \eq\nr{Shard-0} simplifies to 
(denoting now for simplicity $k^{ }_\ini = k$ and 
$k_{\ini\,\parallel} = k_{\parallel}$)
\be
  \la{Shard-0-2nd-line}
  \int \frac{{\rm d} \Omega_{\vec{k}} }{4 \pi} \ 
  \frac{
          k + k_{\parallel} + \eps
  }{
          |\eps|^3
  } \ 
  \theta \big( k + k_{\parallel} + 2\eps \big) 
  \; = \; 
  \frac{
          \theta \big( k + \eps \big) 
  }{
          |\eps|^3
  } \, 
  (k + \eps) 
  \; . 
\ee
We arrive at
\bea
  \Sfunc^{ }_\rmi{hard} (\eps) 
  & = & 
  \frac{g^2}{(2\pi)^2 |\eps|^3} \, 
  \int_{0}^\infty \! {\rm d} k \, 
  k(k + \eps) \, 
  \theta \big( k + \eps \big) \left[ \Phi(k) - \Phi(k+\eps) \right] \, , 
  \la{Shard-1}
\eea
where $\Phi(k)$ was defined in \eq\nr{Phi-def}. 
The above derivation is valid for either sign of $\eps\,$, 
and one may verify that 
$\Sfunc^{ }_\rmi{hard}(-\eps) = -\Sfunc^{ }_\rmi{hard}(\eps)$ 
as expected on general grounds from eq.~\nr{eq:Sofeps}. 
It is thus sufficient to consider $\eps > 0$ below, for which the 
factor $\theta(k+\eps)$ in \eq\nr{Shard-1} can be dropped. 
The remaining integral over $k$ in 
\eq\nr{Shard-1} can be carried out in terms
of polylogarithm functions $\mbox{Li}^{ }_n(z)$. 
We find
\bea
  \Sfunc^{ }_\rmi{hard}(\eps)  
  & \overset{\eps > 0}{=} &
  \frac{ g^2 }{ (2\pi)^2  |\eps|^3} 
  \Big\{\, 
    2 \Nc \Big( \eps T^2  \big[ \lib(0) - \lib(\eps)
   \big] + 2 T^3 \big[ \ltb(0) - \ltb(\eps) \big] \Big)  \nn[2mm] 
   & + &
     \nf \, \sum_{\sigma = \pm} 
   \Big( \, \eps T^2  \big[ \lif(\eps+\sigma\mu) -\lif(\sigma\mu)
   \big] +  2 T^3 \big[ \ltf(\eps+\sigma\mu) - \ltf(\sigma\mu)
   \big] \Big) \, 
   \Big\} 
   \, , \hskip 10mm
   \la{S-eps-hard}
\eea
where 
${l^{ }_\rmi{$n$b}(k) \equiv \mbox{Li}^{ }_n \bigl(e^{-k/T}\bigr) }$ and
${l^{ }_\rmi{$n$f}(k) \equiv \mbox{Li}^{ }_n \bigl(-e^{-k/T}\bigr) }$ 
for $n=2, 3\,$. 
In the first line above, the special values 
$\lib(0)=\tfrac{\pi^2}{6}\,$ and $\ltb(0)=\zeta(3)$ 
are needed.

From here, one can obtain the small and large $\eps$ 
behaviour of $\Sfunc^{ }_\rmi{hard}(\eps)$, 
defined w.r.t.~the hard scale $\Lambda = \max(\pi T,\mu)$, 
namely\footnote{%
  It is not a surprise that the asymptotic limits of 
  $\Sfunc^{ }_\rmii{hard}(\eps)$ 
  below and above the hard scale are identical up to a factor $2\,$, 
  which could be inferred from \eq\nr{Shard-1}. 
  The integral over $k$ must be dominated by $k \sim \Lambda$. 
  When $\eps \gg \Lambda$, 
  the integral in \nr{Shard-1} can thus be approximated by 
  $\eps \int_{0}^\infty \! {\rm d} k \, k  \, \Phi(k)\,$. 
  When $\eps \ll \Lambda$, we can write 
  $\Phi(k) - \Phi(k+\eps) \simeq - \eps\, \Phi'(k)$, 
  and the integral becomes 
  $- \eps \int_{0}^\infty \! {\rm d} k \, k^2 \, \Phi'(k)$, 
  which upon integration by parts is exactly twice the result for 
  $\eps \gg \Lambda\,$.
}
\be
  \la{eq:S_hard_expansions}
  \Sfunc^{ }_\rmi{hard}(\eps)
  \; = \;
  \left\{
  \begin{array}{l}
    \displaystyle \ 
    \frac{\mD^2}{2 \eps^2} + {\cal O}\big( \eps^{-1} \big)
    \\[2mm]
    \displaystyle \ 
    \frac{\mD^2}{4 \eps^2} + {\cal O}\big( \eps^{-3} \big)
  \end{array} \right. 
  \ \mbox{for} \ 
  \left.
  \begin{array}{l}
    \eps \ll \Lambda
    \\[2mm]
    \eps \gg \Lambda
  \end{array} \right. 
  \; ,
\ee
where the Debye mass $\mD$ is given by \eq\nr{eq:mD}. 

%
\subsection{Interpolation between soft and hard domains}
\label{sec:matching}

The expressions in \eqs\nr{S-eps-HTL} and \nr{S-eps-hard} 
are valid when $|\eps| \leq \eps^\star$ and $|\eps| \geq \eps^\star$, 
respectively, 
where we recall that $\eps^\star$ is an arbitrary scale satisfying  
$
  \mD 
  \ll \eps^\star 
  \ll \Lambda  
  = {\rm max}(\pi T, \mu)\,
$. 
Importantly, the behaviour of $\Sfunc^{ }_\rmii{HTL}$ at large $\eps \gg \mD$, 
namely, 
$
  \Sfunc^{ }_\rmii{HTL}(\eps) 
  = 
  \Sfunc^{ }_\rmii{T}(\eps) + \Sfunc^{ }_\rmii{L}(\eps) 
  \simeq 
  \frac{\mD^2}{2\eps^2}
$ 
(cf. appendix~\ref{app:Taylor-ex}), 
coincides with that of $\Sfunc^{ }_\rmi{hard}$ for $\eps \ll \Lambda$ 
given in \eq\nr{eq:S_hard_expansions}. The rate defined by 
\be
\la{rate-match}
  \Sfunc^{ }_\rmi{match}(\eps) 
  \; \equiv \;
  \theta\big(\eps^\star -  |\eps|  \big)  \, \Sfunc^{ }_\rmii{HTL}(\eps) 
  + 
  \theta\big( |\eps| - \eps^\star \big) \, \Sfunc^{ }_\rmi{hard}(\eps)  
  \; , 
\ee
where 
$\Sfunc^{ }_\rmii{HTL}$ is from \eq\nr{S-eps-HTL} 
and 
$\Sfunc^{ }_\rmi{hard}$ is from \eq\nr{S-eps-hard}, 
will thus hold for any $\eps$
and combine the resummed HTL result in the 
soft domain with the hard production rate. 
Moreover, $\Sfunc^{ }_\rmi{match}$ will be independent 
of the choice of $\eps^\star$ as long as 
$\mD \ll \eps^\star \ll \Lambda\,$.  

%
\begin{figure}[t]
\centerline{
  \includegraphics[scale=.65]{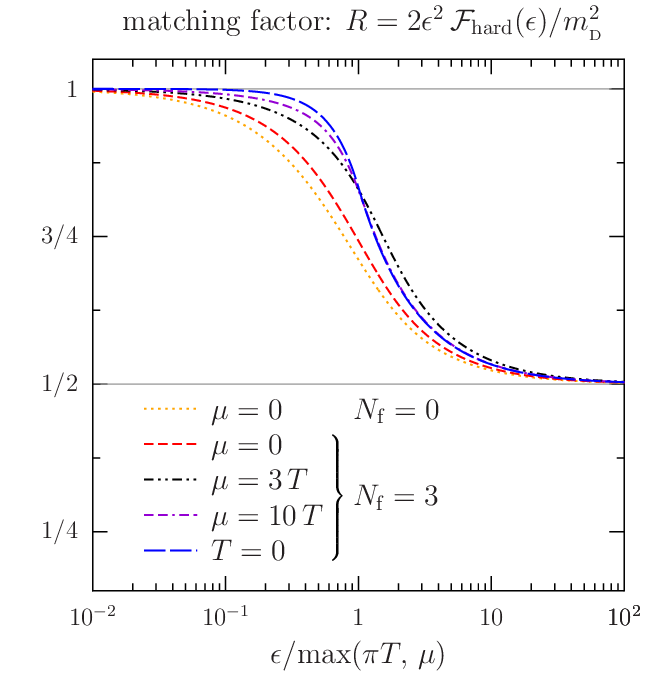}
}
\vspace*{-1mm}
\caption[a]{\small 
  The matching factor  
  $R$, defined in \eq\nr{rate-full}, 
  as a function of $\epsilon$ in units of the hard scale 
  $\Lambda = {\rm max}(\pi T ,\mu)\,$. 
  Several choices of $\nf\,$, $T$ and $\mu$ are shown 
  (in the $\nf=0$ case, obviously $\mu =0$). 
  The limiting values of $1$ and $\frac12$ are 
  indicated by horizontal gray lines. 
}
\la{fig:Rcal}
\end{figure}
%

For practical implementation, we 
use the multiplicative matching defined by 
\be
\la{rate-full}
  \Sfunc^{ }_\rmi{full}(\eps) 
  \; \equiv \; 
  \Sfunc^{ }_\rmii{HTL}(\eps) \, 
  \underbrace{\,
    \frac{\Sfunc^{ }_\rmi{hard}\big(|\eps|\big)}{{\mD^2}/{(2\eps^2)}} \, 
  }_{ \equiv R } \ .
\ee
For $\eps \gg \mD$ we have 
$\Sfunc_\rmi{full} \simeq \Sfunc_\rmi{hard}\,$, 
and for $\eps \ll \Lambda$ (where 
HTL self-energies are needed at $\eps \,\lsim\, \mD$), 
we have 
$\Sfunc_\rmi{full}\simeq \Sfunc_\rmii{HTL}\,$. 
In the overlapping domain 
$\mD \ll \eps \ll \Lambda$, 
obviously 
$
  \Sfunc_\rmi{full}
  \simeq 
  \Sfunc_\rmii{HTL}
  \simeq 
  \Sfunc_\rmi{hard}
  \simeq \frac{\mD^2}{2\eps^2}
  \,
$. 
This shows that in the strict perturbative limit $\mD \ll \Lambda\,$, 
the functions 
$\Sfunc^{ }_\rmi{full}$ 
and 
$\Sfunc^{ }_\rmi{match}$ 
defined by eqs.~\nr{rate-full} 
and \nr{rate-match} coincide. 

It is convenient to introduce $R\,$, 
dubbed a ``hard matching coefficient'' in 
the language of effective field theory, 
as defined in \eq\nr{rate-full}, 
by which $\Sfunc_\rmii{HTL}$ must be multiplied 
to reproduce the correct behaviour when $\eps \gsim \Lambda\,$. 
The ratio $R$ smoothly changes from $1$ to $\frac12$ 
as $\eps$ crosses from below to above the hard scale 
(cf. \eq\nr{eq:S_hard_expansions}), 
and is represented in fig.~\ref{fig:Rcal} 
for different choices of $T$ and $\mu\,$.

Note that in realistic applications, where the soft scale $\mD$ 
and hard scale $\Lambda$ 
are not always well separated in the sense of the strict perturbative limit, 
the expressions \nr{rate-match} and \nr{rate-full} 
do not exactly coincide. 
Since in that case, using 
an ``intermediate scale'' $\eps^\star$ 
as in \eq\nr{rate-match} becomes specious, 
in what follows we will use \eq\nr{rate-full} 
as a {\em definition} of $\Sfunc(\eps)$, 
which doesn't introduce any extra parameter 
and incorporates the correct limiting behaviours for 
both $\eps \,\lsim\, \mD$ and $\eps \,\gsim\, \Lambda\,$. 
Even in situations where the scale separation is 
only moderate, the uncertainty inherited from the matching 
prescription is expected to be minimal. 
This is because the role of $R$ is primarily to correct 
the coefficient of the universal $1/\eps^{2}$ tail, 
starting from $\eps$ values on the order of the hard scale, 
where $\Sfunc_\rmii{HTL}$
is already small (see fig.~\ref{fig:S}). 

%
\section{Evaluation of the quenching weight} 
\la{sec:quenching-weight}

In this section we present our results for $f(x,\Delta)$ 
defined by \eq\nr{eq:bromwich}, 
when the differential scattering rate 
$w(\eps)$ is given by \nr{eq:w-eps-general} 
and 
$\Sfunc(\eps)$ by \nr{rate-full}.
This will be done in sec.~\ref{sec:exact-f} using 
the full expression of $I(\nu)$ defined in~\eq\nr{eq:bromwich}, 
or equivalently 
\eq\nr{eq:Inu-3}. Before that, however, we first discuss 
several interesting features of $f(x,\Delta)$, 
by considering two simplifying limits. 
First, in sec.~\ref{sec:Landau-limit} we verify that for 
a very large path length 
in a certain scaling regime (to be defined below), 
$f(x,\Delta)$ matches with the well-known ``universal'' 
Landau distribution~\cite{Landau:1944fvs}.
Then, in sec.~\ref{sec:T=0} we consider 
the $T \to 0$ limit of $f(x,\Delta)$ at non-zero chemical potential $\mu$.  

The numerical evaluation of $f(x,\Delta)$ 
uses methods detailed in appendix~\ref{app:numerics}, 
and we make our C++ program publicly  available~\cite{zenodo_link}. 

%
\subsection{Landau limit}
\la{sec:Landau-limit}

It is interesting to verify that for a sufficiently large path length $x$
(\ie\ in the formal limit $x \to \infty$), 
and under certain conditions regarding the scaling of $\Delta$ in this limit, 
the distribution \nr{eq:bromwich} tends to the Landau distribution. 
This limit is universal in the sense that it only depends on the 
behaviour of the function $w(\eps)$ for large energies. 
In ref.~\cite{Landau:1944fvs}, 
the incoming particle was considered to undergo a collision with an atom, 
implying $w(\eps) =  \frac{e^4 \, n}{ 8 \pi \, m \, v^2 \eps^2}$ 
(for an incident electron of mass $m$ and velocity $v$, 
and an atomic number density $n$),\footnote{%
  In his original paper~\cite{Landau:1944fvs}, 
  Landau used units of electric charge where the 
  fine structure constant is $\alpha = e^2/(\hbar c)$ 
  instead of $\alpha = e^2/(4\pi \varepsilon_0 \hbar c)\,$. 
  In the present paper, we use natural units and 
  so take $\varepsilon_0 \hbar c = 1\,$.
} 
when $\eps$ is much larger than 
the characteristic binding energy of electrons in the target. 
Here we consider a fast heavy quark crossing 
a QCD plasma characterised by a temperature $T$, 
chemical potential $\mu\,$, and gauge coupling $g$. 
When $\eps \gg \Lambda = \max(\pi T,\mu)$, 
the factor $\nB(\eps)$ in \eq\nr{eq:w-eps-general} can be neglected, 
and using \eq\nr{eq:S_hard_expansions} 
we see that $w(\eps)$ 
also behaves as $\sim 1/\eps^2$ at large $\eps\,$, namely, 
$
  w(\eps) 
  \simeq
  \alpha_s \CF \frac{\mD^2}{4 \eps^2}\,
$. 
As we shall see, this parametric dependence 
is sufficient on its own to obtain the Landau distribution. 

In the following we work out the expression \nr{eq:bromwich}, assuming that 
the contour integral over $\nu$ arises dominantly from the domain 
$\nu \to 0$.\footnote{%
  Here we choose the Bromwich contour ${\cal B}$ 
  (defined after \eq\nr{eq:bromwich}) 
  in the complex $\nu$-plane to be the imaginary axis, \ie\ $\nu_0 = 0$, 
  in order to allow $\nu \to 0$ on the contour. 
} 
After the main result is obtained (see eqs.~\nr{Landau-limit}--\nr{phiL-def} below), 
we will verify that the $\nu \to 0$ region is indeed dominant in 
certain specific domains of the parameters $x$ and $\Delta$, 
justifying a posteriori the applicability of our derivation. 
We therefore need the limit of $I(\nu)$ when $\nu \to 0$. 
In this limit, 
the second term of \nr{eq:Inu-3} is $\sim \rmO(\nu^2)$ 
(using $1- \cosh{\nu\eps} \simeq - \nu^2 \, \eps^2 /2$ 
yields an integral over $\eps$ which is convergent due to 
the presence of the thermal weight $\nB(\eps)$\,), 
which turns out to be negligible compared to the first term of \nr{eq:Inu-3}, 
on which we now focus. 
Introducing an arbitrary scale $\Lambda^\star$, this term can be written as 
\bea
\la{I1-small-nu-split}
  \int_{0}^\infty 
  \dd\eps\, \Sfunc(\eps) \, \big( 1- e^{-\nu\eps} \big) 
  & = &
  \int_{0}^{\Lambda^\star}  \!\! \dd\eps\, 
  \Sfunc(\eps) \, \big( 1- e^{-\nu\eps} \big) 
  + 
  \int_{\Lambda^\star}^\infty  \!\! \dd\eps 
  \left[ \Sfunc(\eps) -  \frac{\mD^2}{4 \eps^2}  \right]
  \big( 1- e^{-\nu\eps} \big) 
  \nn[2mm]
  & + & 
  \int_{\Lambda^\star}^\infty  \!\! \dd\eps\, 
  \frac{\mD^2}{4 \eps^2} \, \big( 1- e^{-\nu\eps} \big)  \ .
\eea
When $\nu \to 0$, 
the limit of the first two terms of the latter expression can be 
obtained by using $1- e^{-\nu\eps} \simeq \nu \, \eps$, 
since the resulting integrals over $\eps$ are convergent, namely, 
\be
  \la{a1a2-def}
  \int_{0}^{\Lambda^\star} \! \dd\eps\, \Sfunc(\eps) \, \eps 
  \; \equiv \;
  a_1 \, \mD^2 
  \, ; 
  \hspace{5mm}
  \int_{\Lambda^\star}^\infty  \dd\eps \,
  \left[ \Sfunc(\eps) -  \frac{\mD^2}{4 \eps^2}  \right] \,
  \eps
  \; \equiv \; 
  a_2 \, \mD^2   
  \, , 
\ee
where the ``constants'' $a_1$ and $a_2$ can be numerically determined for 
any $T$, $\mu$, $\mD$ and $\Lambda^\star$. 
Note that the second integral in \nr{a1a2-def} is convergent 
due to 
$\Sfunc(\eps) = \frac{\mD^2}{4 \eps^2} + {\cal O}\big( \eps^{-3} \big)$ 
when $\eps \to \infty$, see \eq\nr{eq:S_hard_expansions}. 
The first two terms in \eq\nr{I1-small-nu-split} thus contribute to 
$(a_1+a_2) \, \mD^2 \, \nu$ when $\nu \to 0\,$. 
As for the last term of \eq\nr{I1-small-nu-split} 
we have\footnote{\la{foot:I-small-nu}%
  Note that when $\nu \to 0\,$, 
  the logarithmic term in \eq\nr{I1-small-nu-last-term} 
  arises from the logarithmic interval $\Lambda^\star < \eps < 1/\nu\,$, 
  probing a region where $\eps$ can be much larger than $\mu$, 
  $T$ and the hard scale $\Lambda\,$.
}
\be
  \int_{\Lambda^\star}^\infty  \! \dd\eps\; \frac{1- e^{-\nu\eps} }{\eps^2}
  \; = \;  
  \frac{1- e^{-\nu \Lambda^\star}}{\Lambda^\star}
  \, + \,
  \nu\, \Gamma(0,\nu \Lambda^\star) 
  \; \simeq \;
  \nu \big(
    1 \, - \, \gammaE \, - \, \log{\left(  \nu \Lambda^\star \right)} 
  \big)
  \, + \, \rmO(\nu^2) \, ,
  \la{I1-small-nu-last-term}
\ee
where we used 
$
  \Gamma(0,z) 
  \equiv 
  \int_z^{\infty} \dd t \, \frac{e^{-t}}{t} 
  \simeq 
  - \gammaE - \log{z} + \rmO\big( z \big)
$ 
at small $z$, with $\gammaE$ being Euler's constant.
Altogether, $I(\nu)$ exhibits the small $\nu$ limit
\be
\la{I1-small-nu}
  I(\nu) 
  \; \mathop{\ \ \simeq \ \ }_{\nu \to 0} \;  
  \textstyle{\frac14}  \alpha_s \CF \mD^2 \, \nu 
  \bigg[ \,\underbrace{ \, 4(a_1 +a_2) \, +
                        \, 1\,  -\,  \gammaE
                        \, -  
                        \log{\big( \textstyle{\frac{\Lambda^\star}{\mD}}\big)} 
                      }_{ \equiv \, {\displaystyle\kappa} }
  - \, \log{\left( \mD \nu\right)} 
  \, \bigg] \ .
\ee
In this expression the ``constant'' $\kappa$ is independent of 
the choice of $\Lambda^\star$ 
(as can be verified by differentiating it with respect to $\Lambda^\star$, 
the dependence of $a_1$ and $a_2$ 
on $\Lambda^\star$ being given by \eq\nr{a1a2-def}), 
but still depends on $T$, $\mu$ and $\mD$ through $\Sfunc(\eps)$. 
However, since $\kappa$ is dimensionless, 
it is determined by two dimensionless parameters, 
for instance $T/\mD$ and $\mu/\mD\,$.

Inserting \eq\nr{I1-small-nu} into \eq\nr{eq:bromwich} we obtain
\be
  \la{Landau-limit-0}
  f(x,\Delta)  
  \, \simeq \,  
  \int_{\cal B} \,
  \frac{{\rm d} \nu}{2\pi i} 
  \exp 
  \Bigl\{\,
  \nu \Delta \, - \, \bar{x} \, 
  \mD \nu 
  \big[ \kappa - \log{\left( \mD \nu\right)} \big] 
  \, \Bigr\} \ , 
\ee
where we introduced the dimensionless quantity 
\be
\la{xbar-var}
  \bar{x} 
  \; \equiv \; 
  \textstyle{\frac14} \alpha_s \CF \mD \, x \ . 
\ee
Using the variable $u = \bar{x} \, \mD \, \nu$ 
(similarly to ref.~\cite{Landau:1944fvs}) we 
get 
\be
  \la{Landau-limit}
  f(x,\Delta) 
  \, \simeq \, 
  \frac{1}{\mD \,\bar{x}} \ 
  \varphi \left( \lambda \right) \ ; 
          \hspace{6mm} 
          \lambda 
          \; \equiv \; \frac{\Delta}{ \mD \,\bar{x}} 
                - \log{\bar{x}} 
                - \kappa  \ ,
\ee
where the function $\varphi$ is the so-called 
Landau distribution defined by 
\be
  \la{phiL-def}
  \varphi \left( \lambda \right) 
  \; \equiv \; 
  \int_{\cal B} \,
  \frac{\dd u}{2\pi i} \, e^{u \log{u} + \lambda u} \ . 
\ee

We emphasize that \eq\nr{Landau-limit-0}, 
or equivalently eqs.~\nr{Landau-limit}--\nr{phiL-def}, 
have been derived assuming that the $\nu$-integral in \eq\nr{eq:bromwich}
arises dominantly from $\nu \to 0$. 
Therefore, these equations are valid provided the typical value of 
the integration variable $u = \bar{x} \, \mD \, \nu$ in \eq\nr{phiL-def} 
is consistent with the $\nu \to 0$ limit. 
When $\lambda \, \gsim \, {\cal O}(1)$, 
we can infer from \eq\nr{phiL-def} 
that $u \sim {\cal O}(1/\lambda)$,\footnote{%
  This is obvious for $\lambda \sim {\cal O}(1)$.
  For positive $\lambda \gg 1$, 
  it follows from 
  eq.~(24) of ref.~\cite{Landau:1944fvs}.
} 
implying 
$
  1/\nu \sim \lambda \,\mD\,\bar{x} 
        \sim \Delta -  \mD \,\bar{x} \log{\bar{x}}
  \,
$. 
We thus observe that the $\nu \to 0$ limit is consistent with 
two regions of the parameters $x$ and $\Delta\,$, 
namely, the region of large $x$ at fixed $\lambda \sim {\cal O}(1)$ 
(the condition $\lambda \sim {\cal O}(1)$ being referred to 
as ``Landau scaling''), 
and the region of large (positive) $\lambda$ at fixed $x$ 
(which is equivalent to large $\Delta$ at fixed $x$). 
Therefore, eq.~\nr{Landau-limit} 
is fully valid in the two latter parametric limits.
This will be clearly illustrated by fig.~\ref{fig:T=0} (right panel) in the following section.  

Finally, let us note 
that the appearance of the Landau distribution 
in these parametric domains 
is a direct consequence of 
the asymptotic behaviour $w(\eps) \sim 1/\eps^2$ at $\eps \gg \Lambda$ 
(leading to the logarithmic term in \eq\nr{Landau-limit-0}), 
regardless of the exact form of $w(\eps)$ 
for $\eps \,\lsim \, \Lambda\,$. 
(The latter range 
only affects the constant $\kappa$ appearing 
in \eq\nr{Landau-limit}.)

%
\subsection{Finite path length: cold dense medium}
\la{sec:T=0}

We verified in the previous section that
for $x \to \infty$ in the Landau scaling regime $\lambda \sim {\cal O}(1)$, 
$f(x,\Delta)$ defined by \eq\nr{eq:bromwich} approaches 
the Landau distribution in \eq\nr{Landau-limit}, 
regardless of the parameters $T$, $\mu$ and $\mD$ which determine 
$w(\eps)$. In this section we consider $f(x,\Delta)$ for 
{\em finite $x$} and a cold dense medium, \ie, 
taking the limit $T \to 0$ at fixed $\mu \neq 0\,$. 
Only the first term of \eq\nr{eq:Inu-3} contributes in this situation.
Thus, this case deals with deviations of $f(x,\Delta)$ 
from the Landau distribution, which are due solely to keeping $x$ finite. 

For $T = 0$, we denote the distribution by $f_{_{T=0}}(x,\Delta)\,$, 
which depends on the scales $\mu$ and $\mD \sim g \mu$ (cf. \eq\nr{eq:mD}) 
contained in $\Sfunc(\eps)$, 
for any finite $x$. From the previous section, 
it is clear that $f_{_{T=0}}(x,\Delta)$ will tend towards 
the Landau distribution when $x \to \infty$ 
in the Landau scaling regime (as for any $T$, $\mu$ and $\mD$). 
Note that in the absence of thermal fluctuations, 
$1+\nB(\eps) \to \theta(\eps)$ and the rate~\nr{eq:w-eps-general} 
corresponds to energy loss only. As a result $f_{_{T=0}}(x,\Delta)$ 
must similarly correspond to only energy loss ($\Delta \geq 0$).\footnote{%
  This follows from \nr{eq:bromwich}. 
  When $\Delta < 0$, the Bromwich integration contour can be closed in 
  the right half of the complex $\nu$ plane, 
  where $I(\nu)$ has no singularity (for $T=0$), 
  leading to $f_{_{T=0}}(x,\Delta<0)=0$. 
  (The distribution $f(x,\Delta)$ will acquire a support for $\Delta < 0$ 
  as soon as $T \neq 0$, see sec.~\ref{sec:exact-f}.)
} 
The situation considered here is analogous to the case of a test charge 
passing through normal, cold matter, 
the scale $g\mu$ playing the role of an infrared atomic scale. 

Let us mention that when $T =0$ and $\mu \neq 0$, 
$\Sfunc_\rmi{hard}(\eps)$ takes a simple form, 
which can be obtained by taking the $T \to 0$ limit of \eq\nr{S-eps-hard}, 
or more conveniently using \eq\nr{Shard-1} 
with $\Phi(k) =  \nf \, \theta(\mu -k)$ 
(following from \eq\nr{Phi-def} when $T \to 0$). 
For $\eps >0$ (the only case which matters here, as mentioned above), 
an elementary calculation gives: 
\be
  \la{S-hard-zeroT}
  \left. \Sfunc_\rmi{hard}(\eps) \right|_{T=0} 
  \; \overset{\eps > 0}{=} \; 
  \frac{\mD^2}{2\eps^2} 
  \underbrace{\,\left[\, 
    \theta( \mu - \eps) \left( 1- \frac{\eps^2}{6 \mu^2} \right)
    + 
    \theta( \eps - \mu) \left( \frac{1}{2} + \frac{\mu}{3 \eps} \right) 
              \,\right]\,}_{ = \, R } \ . 
\ee
Thus, for $T=0$ the ratio ${R}$ defined in \eq\nr{rate-full} is 
a simple piece-wise function of $\eps/\mu$. 
This function and its derivative are 
continuous at the point $\eps = \mu\,$. 
We also observe the change of behaviour of $\Sfunc_\rmi{hard}(\eps)$ 
when crossing the hard scale ($\Lambda=\mu$ in the present case), 
from ${\mD^2}/{(2\eps^2)}$ to ${\mD^2}/{(4\eps^2)}$, 
as is the case for any $T$ and $\mu$, 
see \eq\nr{eq:S_hard_expansions}.

An important feature of the $T =0$ case is that 
the total scattering rate \nr{damping-rate}, namely, 
\be
  \la{total-rate-vac}
  \Rate_{_{T=0}} 
  \; \equiv \;  
  \Rate^{ }_{0} 
  \; = \; 
  \alpha_s \CF  \int_0^{\infty} \! \dd \eps 
  \,\, \Sfunc_\rmi{full}(\eps) 
  \; ,
\ee
is finite, the above integral over $\eps$ being convergent at $\eps \to \infty$ 
(where $\Sfunc_\rmi{full}(\eps) \sim 1/\eps^2$) 
but also at $\eps \to 0$ 
(where $\Sfunc_\rmi{full}(\eps) \simeq \Sfunc_\rmii{HTL}(\eps) \simeq \frac13\,$, 
cf. eqs.~\nr{ST_expansions}--\nr{SL_expansions}). 
Thus, $f_{_{T=0}}(x,\Delta)$ has a ``no-scattering'' component, 
proportional to $\delta(\Delta)$,
which can be singled out as in \eq\nr{f-delta-piece}. 
The total rate $\Rate^{ }_{0}$ 
depends on the two parameters $\mu$ and $\mD$ via $\Sfunc_\rmi{full}(\eps)$, 
and can be evaluated numerically for any values of these parameters.\footnote{%
  Note that the $\eps$-integral in \nr{total-rate-vac} 
  is dominated by $\eps \sim  \rmO(\mD)\,$. 
  Thus, in the perturbative limit $\mD \ll \mu\,$, 
  the integral depends negligibly on $\mu$ and 
  the rate $\Rate^{ }_{0}$ is a function of $\mD$ only.
}

%
\begin{figure}[t]

\centerline{
    \includegraphics[scale=.65]{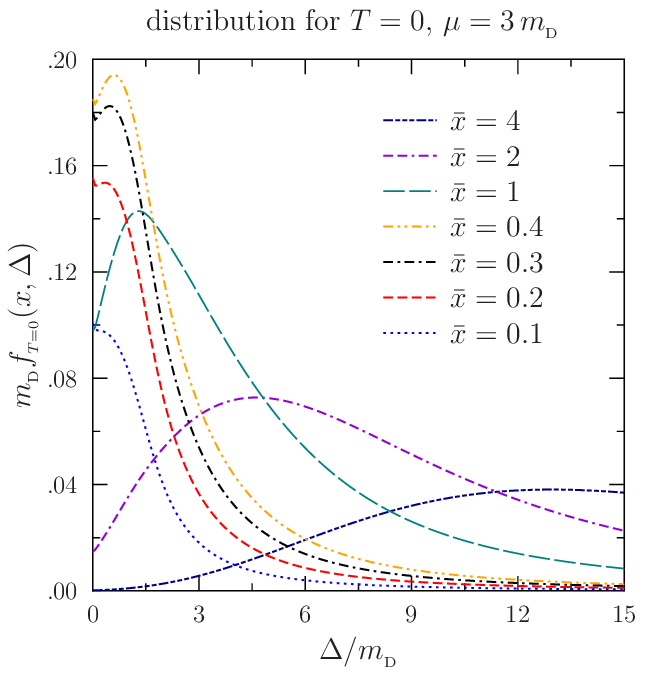}
  ~~\includegraphics[scale=.65]{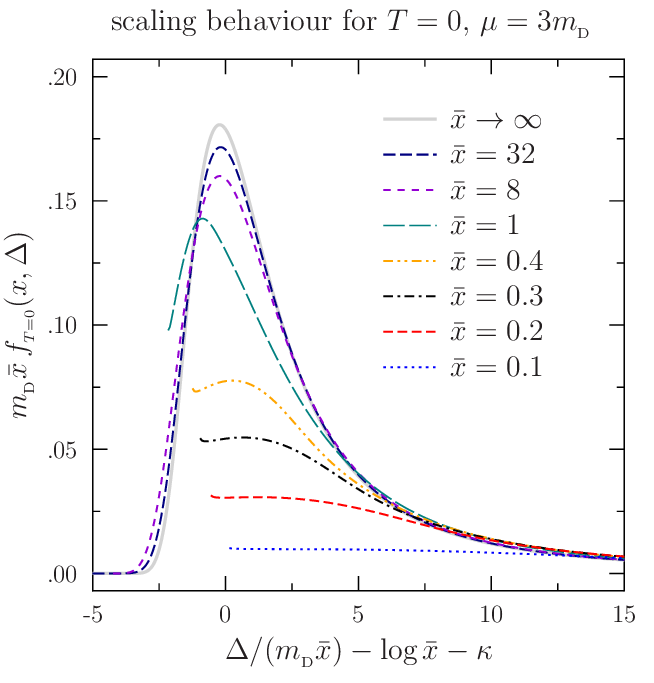}
}

\vspace*{1mm}

\caption[a]{\small 
  Energy loss distribution $f^{ }_{_{T=0}}(x,\Delta)$ 
  with $\mu = 3 \mD\,$, for several values of  
  the dimensionless variable $\bar x$ (defined in \nr{xbar-var}). 
  The left panel plots the distribution as a function of $\Delta\,$, 
  in units of $\mD\,$. 
  The same curves are shown in the right panel in terms of 
  the 
  Landau scaling variable $\lambda$ 
  appearing in \eq\nr{Landau-limit}
  (here $\kappa$ defined in \eq\nr{I1-small-nu} is $\kappa \approx 2.16$),
  better illustrating how $f^{ }_{_{T=0}}(x,\Delta)$ 
  approaches the Landau distribution 
  when $\bar x \to \infty$ 
  at fixed $\lambda \sim {\cal O}(1)$. 
  We note that $f_{_{T=0}}$ is strictly zero for $\Delta < 0\,$, 
  thus having a discontinuity which is visible in the right panel. 
  The distribution also contains a Dirac-$\delta$ component (not plotted) 
  as discussed in the main text. 
  }
\la{fig:T=0}
\end{figure}
%

In fig.~\ref{fig:T=0} we display $f_{_{T=0}}(x,\Delta)$ 
as a function of $\Delta$ for different values of $x\,$,
to illustrate how the distribution approaches 
the Landau limit \nr{Landau-limit} when $x$ increases, 
for $\mu = 3\mD$ (corresponding to $g \approx 0.86$ for $\nf=3$). 
Note that the functions shown  
correspond to $f_{_{T=0}}(x,\Delta)$ 
not including the Dirac-$\delta$ term (first term of \eq\nr{f-delta-piece}). 
These functions are thus not exactly normalised.\footnote{%
  For each value of $x$, the contributions to the integral 
  $\int_0^{\infty} \dd \Delta \, f_{_{T=0}}(x,\Delta)$ of 
  the displayed function and of its associated 
  $\delta(\Delta)$-term are $1-e^{- x  \Rate^{ }_{0}}$ 
  and $e^{- x  \Rate_{0}}$, respectively. 
  Only the sum of the displayed function and 
  the $\delta(\Delta)$-term is 
  a properly normalised probability distribution.
} 

Some details concerning the numerical evaluation 
of $f_{_{T=0}}(x,\Delta)$ are given in appendix~\ref{app:numerics}. 
It is worth mentioning that, according to \eq\nr{vac_sol1}, 
the distribution $f_{_{T=0}}(x,\Delta)$ can be expressed 
in a form similar to \eq\nr{Landau-limit} as\footnote{%
  Here we omit the implicit dependence on 
  the ratio $\mu/\mD\,$, 
  which is akin the coupling $g\,$. 
  }
\be
  \la{scaling-vac}
  f^{ }_{_{T=0}}(x,\Delta)
  \; = \; 
  \frac{1}{\mD \, {\bar{x}}} 
  \, 
  \phi^{ }_{_{T=0}} \bigg( 
      \,  \bar x  \, 
     ,\, \frac{\Delta}{\mD }\, 
  \bigg) \, , 
\ee
where $\phi^{ }_{_{T=0}}$ is a function 
which, for compatibility with \eq\nr{Landau-limit}, satisfies 
\be
 \la{phi-T=0-limit}
  \lim_{\bar x \to \infty} 
  \phi^{ }_{_{T=0}} \bigg( 
      \,  \bar x  \, 
     ,\, \frac{\Delta}{\mD }\, 
  \bigg) 
  \; = \;  
  \varphi \left( 
  \frac{\Delta}{ \mD \,\bar{x}} - \log{\bar{x}} - \kappa 
  \right)  \ . 
\ee
When $\bar x \to \infty$ the dependence on the two arguments of  
$f^{ }_{_{T=0}}$ collapses into 
a dependence on the particular combination 
$
  \frac{\Delta}{\mD \bar x} 
  - \log{\bar x}\,
$. 
This scaling in a single variable is also found in the limit of 
large $\Delta$ for fixed $\bar x$, 
where the distribution also tends towards the Landau distribution 
(as emphasised below \eq\nr{phiL-def}). 
These features are illustrated in fig.~\ref{fig:T=0} (right). 

%
\begin{figure}[t]
\centerline{
  \includegraphics[scale=.65]{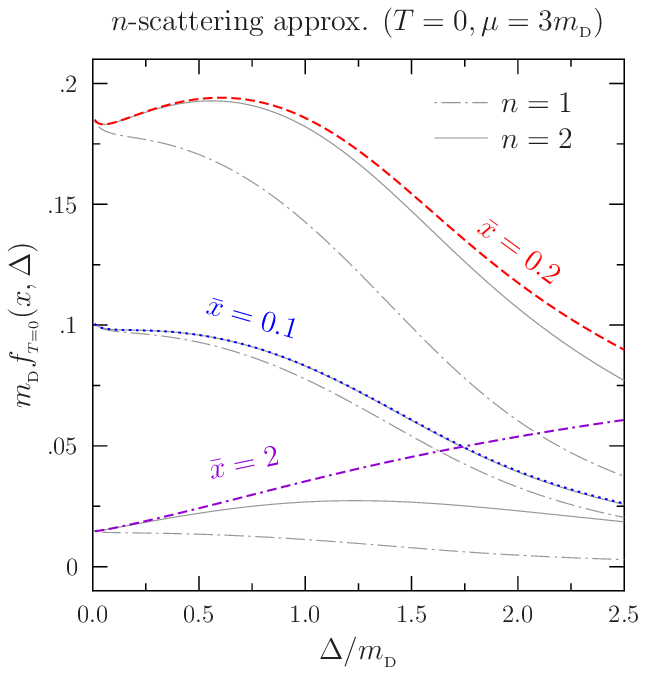}
}
\caption[a]{\small
  Magnification of the small $\Delta$ behaviour of $f^{ }_{_{T=0}}(x,\Delta)$ 
  (for $\mu = 3 \mD$) from the left panel of fig.~\ref{fig:T=0}, 
  showing the cases $\bar x = \{ 0.1, 0.2, 2 \}$. 
  The thin gray lines are from the two lowest order terms shown in 
  \eq\nr{small-Delta-approx} (not counting the Dirac-$\delta$). 
  The accuracy of the ``two-scattering'' truncated 
  expansion obviously deteriorates when $\Delta$ or $\bar x$ grows. 
}
\la{fig:small-Delta}
\end{figure}
%

Let us mention that at sufficiently small $\Delta\,$, 
where $f_{_{T=0}}(x,\Delta)$ strongly deviates from 
the Landau distribution (if $x$ is not too large), 
the energy loss should come from only a few scatterings. 
This suggests that $f_{_{T=0}}(x,\Delta)$ should be well-approximated 
by the first terms of the Poisson expansion \nr{f-Poisson}, namely 
\be
  \la{small-Delta-approx}
  f^{ }_{_{T=0}}(x,\Delta) \; \simeq \; e_{ }^{- x \Rate_{0} } 
  \bigg\{\,
    \delta( \Delta )
    \, + \,
    \alpha_s \CF\, x \, \Sfunc(\Delta)
    \, + \,
    \frac{(\alpha_s \CF\,x)^2 }2 
    \int_0^\Delta \dd \eps \, \Sfunc(\eps) \Sfunc(\Delta - \eps)
    \, + \,
    \ldots
  \,\bigg\} \ .
\ee
We observe that after subtracting the $\delta(\Delta)$-term, 
the distribution approaches the finite limit 
$\frac13 \alpha_s \CF\, x e^{- x \Rate_{0}}$ when $\Delta \to 0^+$, 
whereas $f_{_{T=0}}(x,\Delta)=0$ for $\Delta<0\,$. 
This discontinuity at $\Delta = 0$ is clearly visible in fig.~\ref{fig:T=0}. 
The small hook shape visible in fig.~\ref{fig:T=0} for $\bar x \lsim 0.5$ 
at very small $\Delta >0$ can be simply explained as follows. 
When $\Delta$ starts to deviate from zero, 
the single scattering contribution 
(second term $\sim \Sfunc(\Delta)$ of \eq\nr{small-Delta-approx}) 
dominates and decreases with growing $\Delta$ 
(as specified by eqs.~\nr{ST_expansions}--\nr{SL_expansions}). 
This decrease is quickly compensated by the contribution 
of additional scatterings 
(starting with the double scattering given by 
the third term of \eq\nr{small-Delta-approx}), 
which will eventually build the broad peak of the 
distribution.
This is illustrated in fig.~\ref{fig:small-Delta}, 
where we compare $f^{ }_{_{T=0}}(x,\Delta)$ with the 
``two-scattering approximation'' 
obtained by retaining only the second and third terms 
of \eq\nr{small-Delta-approx}, for several values of $\bar x$. 
Note that a similar situation, 
where a peak arises from multiple soft scatterings 
and a tail from a single hard collision, 
occurs in other contexts where $\Rate_{0}$ is finite~\cite{Borghini:2006xh}. 

To conclude this section, let us note the apparent presence of 
two main regimes in the evolution of $f_{_{T=0}}(x,\Delta)$ with $x$. 
For small values of $x$, the initial $\delta(\Delta)$ distribution 
``feeds'' into $f_{_{T=0}}(x,\Delta)$ 
(see the left panel of fig.~\ref{fig:T=0}), 
building its normalization until $x$ reaches a scale 
$\sim 1/\Rate_{0} \sim (\alpha_s \CF \mD)^{-1}$, 
which is on the order of the heavy quark mean free path in 
the cold dense medium~\cite{Vanderheyden:1996bw}. 
As $x$ increases further, 
the distribution quickly adheres to the Landau distribution 
(which is more easily seen in the right panel of fig.~\ref{fig:T=0}). 

%
\subsection{Finite path length: thermal medium}
\la{sec:exact-f}

We now consider the case of a finite path length with $T>0\,$, 
which introduces thermal fluctuations 
into the discussion of the previous section. 
We will numerically evaluate the distribution $f(x,\Delta)$ 
over a broad range of $x$ and $\Delta$ values, 
using \eq\nr{eq:bromwich} and the full expression for $I(\nu)$ 
in \eq\nr{eq:Inu-3}, \ie\ retaining terms proportional to $\nB\,$. 
The comparison between $f(x,\Delta)$ and $f_{_{T=0}}(x,\Delta)$ 
studied in sec.~\ref{sec:T=0} will reveal the role of thermal effects. 
For a given energy loss $\Delta\,$, 
we expect thermal effects to smear the distribution $f_{_{T=0}}(x,\Delta)$ 
around $\Delta$ by some amount $\sim \rmO(T)$. 
In particular, when $T > 0$, energy gain ($\Delta <0$) becomes possible
due to the absorption of thermal gluons by the fast particle.
These smearing and energy gain effects will of course be suppressed 
as $x$ increases (at fixed $\lambda \sim {\cal O}(1)$), 
or $\Delta$ increases (at fixed $x$), 
because in these two limits, 
$f(x,\Delta)$ must approach the Landau distribution, 
as was already the case for $f_{_{T=0}}(x,\Delta)$ (see sec.~\ref{sec:T=0}). 

On the other hand, we expect thermal effects to strongly influence 
the shape of $f(x,\Delta)$ when $\Delta$ and $x$ are both sufficiently small. 
To explore this, we derive the limiting behaviour 
of $f(x,\Delta)$ as $|\Delta| \to 0\,$. 
Starting from \eq\nr{eq:bromwich} and 
parametrising the integration contour as $\nu = i \xi /|\Delta|$ we obtain
\be
  f(x,\Delta) 
  \; = \; 
  \frac{1}{|\Delta|} \int_{-\infty}^{\infty} \, \frac{\dd \xi}{2\pi} \;
  \exp 
  \biggl\{ 
    i \xi \, {\rm sgn}(\Delta) 
    - x \, I \! \left( i \frac{\xi}{|\Delta|} \right) 
  \biggr\} \ .
  \la{eq:bromwich-small-Delta} 
\ee
The $|\Delta| \to 0$ limit of $f(x,\Delta)$ is thus 
related to the behaviour of $I(\nu)$ as $|\nu| \to \infty\,$, 
under conditions that will become apparent in the following calculation. 

To obtain the $|\nu| \to \infty$ limit of $I(\nu)$, we first rewrite 
one of the integrals appearing on the right hand side 
of \eq\nr{eq:Inu-3} as:
\bea
  2 \int_{0}^\infty \dd\eps\, 
  \Sfunc(\eps) \, 
  \nB\big(\eps \big)\, 
  \big(1- \cosh{\nu\eps} \big) 
  & = &
  2 \int_{0}^\infty \dd\eps\, 
  \left[\Sfunc(\eps) -\frac{1}{3} \right] \, \nB\big(\eps \big) 
  \big(1- \cosh{\nu\eps} \big) 
  \nn[2mm]
  & + &
  \frac{2}{3} \int_{0}^\infty \dd\eps\, \nB\big(\eps \big) 
  \big(1- \cosh{\nu\eps} \big) 
  \; , 
  \la{eq:large-nu-exp-1}
\eea
where $\Sfunc = \Sfunc_\rmi{full}$ from \eq\nr{rate-full}. 
When $|\nu| \to \infty$, 
the rapidly oscillating factors $e^{\pm i |\nu| \eps}$ 
in \eq\nr{eq:large-nu-exp-1} can be dropped, except in the last term 
where doing so would produce an infrared divergent integral, 
stemming from $\nB(\eps) \simeq T/\eps$ when $\eps \to 0$. 
The behaviour of the last term of \eq\nr{eq:large-nu-exp-1} 
when $|\nu| \to \infty$ is found using the 
identity~\cite{GradshteynRyzhik}
\be
  \int_0^\infty \dd t \, 
  \frac{1-\cos{A t}}{e^{t}-1} 
  \; \mathop{\ \ \simeq \ \ }_{A \to \infty} \;
  \log{A} + \gammaE + \rmO(A^{-2}) \ . 
\ee
We thus find, neglecting terms which vanish when $|\nu| \to \infty\,$, 
\be
  \la{eq:Inu-large-nu-2}
  I(\nu) \; \mathop{\ \ \simeq \ \ }_{|\nu| \to \infty} \;
  \alpha_s \CF \, \Big\{ \, 
  K \left( T , \mu , \mD \right) 
  + 
  \tfrac{2}{3} \, T \log{ \big( \, |\nu|  \, T  \big) }  
  \, \Big\} \ ,
\ee
where the function $K$ is defined by
\be
  K \left( T, \mu , \mD  \right) 
  \; = \;  
  \frac{2}{3} \gammaE \, T 
   + 
  \int_{0}^\infty \! \dd\eps\, \Sfunc(\eps) 
   + 
  2 \int_{0}^\infty \! \dd\eps 
  \left[\Sfunc(\eps) -\frac{1}{3} \right] \nB\big(\eps \big)  
  \; . 
  \la{eq:Kdef}
\ee

Using the asymptotic behaviour \nr{eq:Inu-large-nu-2} 
of $I(\nu)$ in \eq\nr{eq:bromwich-small-Delta} we obtain
\be
  f(x,\Delta) 
  \; \mathop{\ \ \simeq \ \ }_{|\Delta| \to 0} \; 
  J\left( \frac23 x \alpha_s \CF T \right) \, 
  \frac{
      e^{- x \alpha_s \CF K \left( T , \mu , \mD \right)}
  }{ |\Delta| } \, 
  \left( \frac{|\Delta|}{T} \right)^{ \frac23 x \alpha_s \CF T}  
  \; ,
  \la{eq:f-small-Delta} 
\ee
where 
$
  J(a) 
  \equiv 
  \int_{0}^{\infty} \frac{\dd \xi}{\pi} 
  \frac{\cos{\xi}}{\xi^a}
  =
  \big[{2 \Gamma(a) \sin{\frac{(1-a)\pi}{2}}}\big]^{-1}
$. 
The expression in \eq\nr{eq:f-small-Delta}  
is well-defined provided the $\xi$-integral 
defining $J$ in the first factor is convergent, requiring 
\be
  0 \, < \, \frac{2}{3}x \alpha_s \CF T  \, < \, 1 
  \hspace{5mm} \Longleftrightarrow \hspace{5mm}
  0 \, < \,  {\bar x} \,  < \, \frac{3}{8} \, \frac{\mD}{T} 
  \; ,
  \la{small-Delta-condition}
\ee
with ${\bar x}$ defined in \eq\nr{xbar-var}. 
Thus, when the condition \nr{small-Delta-condition} holds, 
$f(x,\Delta)$ 
given by \eq\nr{eq:f-small-Delta} 
behaves algebraically as
$f(x,\Delta) \sim |\Delta|^{-1+\frac{2}{3}x \alpha_s \CF T}\,$,
whereby it is singular at $\Delta = 0\,$.\footnote{%
  Let us note that beyond the HTL approximation 
  and with magnetic screening at the ultrasoft scale $g^2 T$, 
  we expect the algebraic behaviour \nr{eq:f-small-Delta} 
  to be valid only when 
  the logarithmic interval in $\eps$ contributing to 
  the logarithm in \nr{eq:Inu-large-nu-2}, namely, 
  $
    T \gg \eps \gg  
    \frac{1}{|\nu|} 
    \sim |\Delta|
  $, 
  is unaffected by the ultrasoft scale, \ie\ when $|\Delta| \gsim g^2 T$. 
}

On the other hand, when $\frac{2}{3}x \alpha_s \CF T > 1\,$, 
\eq\nr{eq:f-small-Delta} becomes invalid 
and in that case $f(x,\Delta)$ is finite at $\Delta = 0\,$. 
Indeed, when $\Delta = 0$ the $\nu$-integral in \nr{eq:bromwich} 
converges as $|\nu|^{-\frac{2}{3}x \alpha_s \CF T}$ at large $\nu$, 
by virtue of \nr{eq:Inu-large-nu-2}. 
The transition between the two regimes 
$\frac{2}{3}x \alpha_s \CF T < 1$ and $\frac{2}{3}x \alpha_s \CF T > 1\,$, 
where $f(x,\Delta)$ is respectively singular or finite when $\Delta \to  0\,$, 
is visible in fig.~\ref{fig:small-T} 
where $T$ increases at fixed $x\,$.\footnote{%
  At the transition point 
  $\frac{2}{3}x \alpha_s \CF T = 1\,$, 
  setting $\Delta = 0$ in \eq\nr{eq:bromwich} 
  yields a $\nu$-integral which {\it diverges} 
  as $1/|\nu|$ at large $\nu$. 
  For $\frac{2}{3}x \alpha_s \CF T = 1$, 
  we thus expect $f(x,\Delta)$ to diverge logarithmically 
  when $\Delta \to  0$ 
  (instead of algebraically as for $\frac{2}{3}x \alpha_s \CF T < 1\,$). 
  To prove this, we can still start from \eq\nr{eq:bromwich-small-Delta}, 
  but should notice that when $\frac{2}{3}x \alpha_s \CF T = 1$, 
  using the large $\nu$ expansion \nr{eq:Inu-large-nu-2}  
  leads to a spurious singularity at $\nu \to 0\,$. 
  To remedy this problem, we can replace 
  $
    \log{ \big( \, |\nu| T  \big) }  
    \to 
    \log{ \big( 1+ \, |\nu| T  \big) }
  $ 
  in \eq\nr{eq:Inu-large-nu-2}. 
  This does not affect the validity of the large $\nu$ expansion of $I(\nu)$, 
  and avoids introducing  a spurious singularity at $\nu \to 0\,$. 
  It is then easy to find that for the special case 
  $\frac{2}{3}x \alpha_s \CF T = 1$, 
  the result \nr{eq:f-small-Delta}  should be replaced by 
  \vspace{-1mm}
  \be
    \frac{2}{3}x \alpha_s \CF T \; = \; 1 
    \quad \Longrightarrow  \quad
    f(x,\Delta) 
    \; \mathop{\ \ \simeq \ \ }_{|\Delta| \to 0}\;
    \frac{e^{- x \alpha_s \CF K \left( T, \mu, \mD \right)}}{\pi \, T} 
    \ \log{\left( \frac{T}{|\Delta|} \right)}  \ . 
    \nonumber 
  \ee
  \vspace{-1mm}
}
\begin{figure}[t]
\centerline{
    \includegraphics[scale=.65]{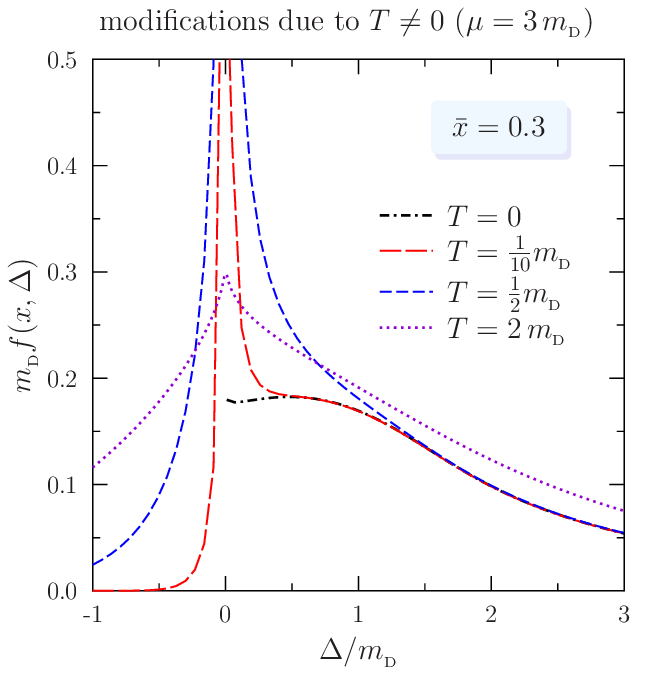}
  ~~\includegraphics[scale=.65]{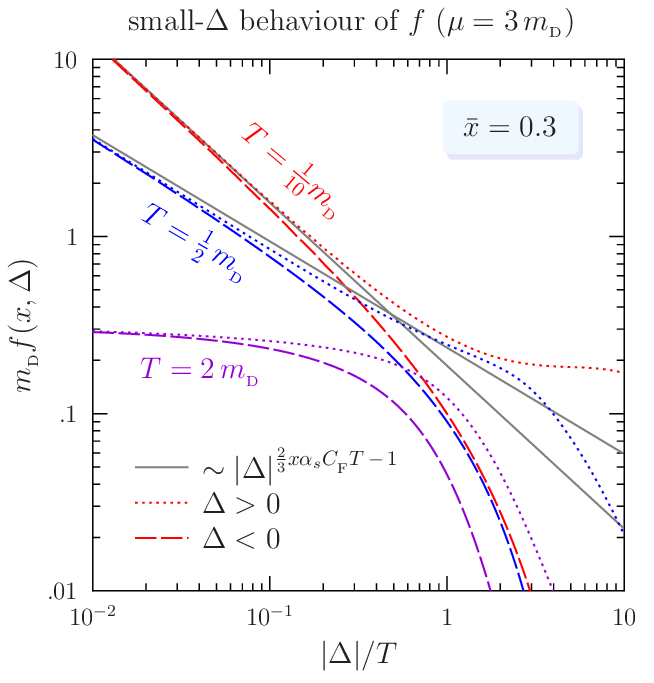}
}
\vspace*{-3mm}
\caption[a]{\small 
  Illustration of the qualitative change in the distribution $f$ 
  when introducing thermal effects. 
  The left panel shows the same $T=0$ curve as in fig.~\ref{fig:T=0} 
  for ${\bar x}=0.3\,$, together with the distributions $f(x,\Delta)$ for small, 
  increasing values of $T$ (all for $\mu = 3 \mD$). 
  The latter distributions exhibit a pronounced peak near $\Delta \simeq 0\,$, 
  which is reminiscent of the Dirac-$\delta$ term in the $T=0$ distribution. 
  The log-log plot in the right panel provides a  
  better visualisation of the singular power-law  behaviour of 
  $f(x,\Delta)$ when $\Delta \to  0\,$, predicted under the condition 
  \nr{small-Delta-condition}. 
  For $\bar x = 0.3\,$, the transition
  from a singular to a finite behaviour when $\Delta \to  0$  
  occurs at the critical value $T = 1.25 \, \mD\,$. 
  The coloured curves show the numerical distributions, while 
  the solid gray lines are the small-$|\Delta|$ 
  asymptotic shape predicted from eq.~\nr{eq:f-small-Delta-small-T}, 
  shown for the two temperatures where it 
  applies, namely $T=\frac{1}{10} \mD$ and  $T = \frac12 \mD\,$. 
}
\la{fig:small-T}
\end{figure}
%

Before proceeding, let us consider how \eq\nr{eq:f-small-Delta} 
simplifies when $T \to 0$ at fixed $\mu\,$, 
corresponding to a situation where we would start from a cold dense medium 
(cf. sec.~\ref{sec:T=0}),  
and switch on a tiny temperature. 
The formula \nr{eq:f-small-Delta} then becomes\footnote{%
  Taking the limit $T \to 0$ at fixed $\mu$ in \eq\nr{eq:Kdef}, 
  we obtain 
  $
    K \! \left( T, \mu, \mD \right) 
    \to 
    \int_{0}^\infty \dd\eps 
    \left. \Sfunc_\rmi{full}(\eps) \right|_{T = 0} 
    = 
    \Rate_{0}/(\alpha_s \CF ) \,
  $ 
  (cf. \eq\nr{total-rate-vac}). 
  We also use the $a \to 0$ limit of $J(a)$ 
  defined below \eq\nr{eq:f-small-Delta}, 
  $J(a) \mathop{\ \simeq \ } \; \frac{a}{2} + \rmO(a^2)$.
}
\be
  f(x,\Delta) 
  \; \mathop{\ \ \ \simeq \ \ \ }_{_{|\Delta| \to 0}}^{^{T \ll |\mu|}} \;
  \frac{e^{- x \Rate_{0}}}{2 |\Delta|} 
  \cdot 
  \frac{2}{3}x \alpha_s \CF T 
  \cdot
  \left( \frac{|\Delta|}{T} \right)^{^{\! \! \frac{2}{3}x \alpha_s \CF T}} 
  \; = \; 
  \frac{e^{- x \Rate_{0}}}{2} 
  \cdot 
  \frac{\partial}{\partial |\Delta|} 
  \left( \frac{|\Delta|}{T} \right)^{^{\! \!\frac{2}{3}x \alpha_s \CF T}} \ .
  \la{eq:f-small-Delta-small-T} 
\ee
Although the latter behaviour is obtained assuming $|\Delta| \ll \rmO(T)$, 
it is noteworthy that integrating \eq\nr{eq:f-small-Delta-small-T} 
over $\Delta$ 
on the interval $- T \leq \Delta \leq T$ equals $e^{- x \Rate_{0}}$. 
In other words, the distribution \nr{eq:f-small-Delta-small-T} 
contributes to the normalization of $f(x,\Delta)$ on this 
interval similarly to how 
the $\delta(\Delta)$-term contributed to the normalization 
of $f^{ }_{_{T=0}}(x,\Delta)$ for $T=0$, 
cf. \eq\nr{small-Delta-approx}. 
This is consistent with the interpretation of \eq\nr{eq:f-small-Delta-small-T} 
as the smearing of the zero-scattering component in the $T=0$ case, 
due to thermal fluctuations. 
Those features are illustrated in fig.~\ref{fig:small-T}, 
for a given value of ${\bar x}$. 
The smearing effect is illustrated in the left panel 
for increasing values of $T$. 
This plot is also consistent with the existence of 
the critical value $T = \frac{3 \mD}{8 {\bar x}}$ below which $f(x,\Delta)$ 
is singular at $\Delta = 0$ (see \eq\nr{small-Delta-condition}). 
When $T < \frac{3 \mD}{8 {\bar x}}$, the 
agreement with the small-$|\Delta|$ asymptotic prediction 
is shown in the right panel of fig.~\ref{fig:small-T}.
\begin{figure}[t]
\centerline{
    \includegraphics[scale=.65]{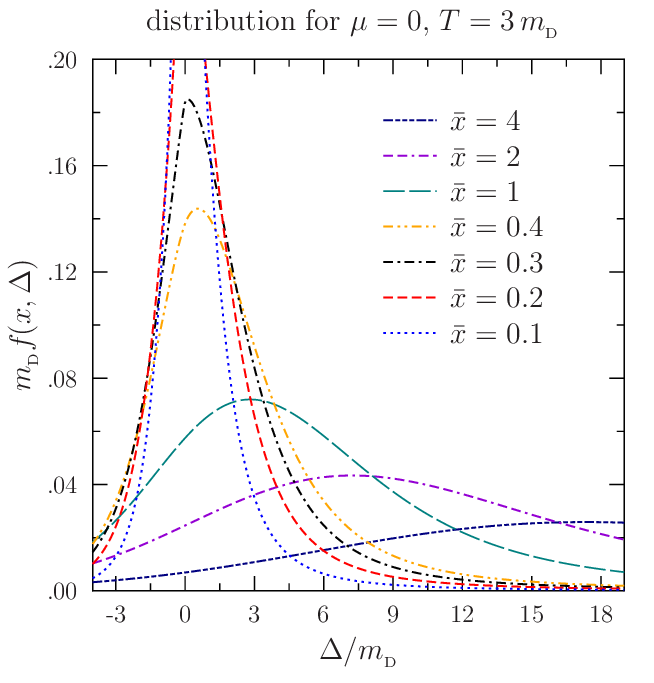}
  ~~\includegraphics[scale=.65]{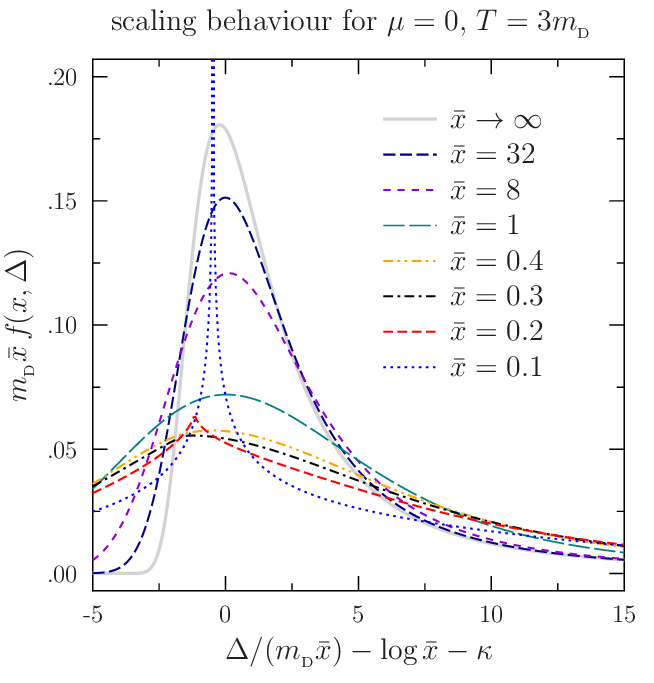}
}
\vspace*{-2mm}
\caption[a]{\small 
  Results for the full distribution, demonstrating the
  feature of energy gain ($\Delta < 0$). 
  The same $\bar x$ values are shown in fig.~\ref{fig:T=0} for $T=0$ 
  and $\mu = 3 \, \mD\,$. 
  Here $T = 3 \, \mD\,$, $\mu =0$ and $\nf = 3\,$. 
  The left panel shows the distribution in units of $\mD$ 
  and the right panel shows the same curves  
  scaled as in \eq\nr{Landau-limit} 
  (with $\kappa \approx 2.78$ in the present case). 
  We see how the initial distribution $f(x=0,\Delta) = \delta(\Delta)$ 
  broadens when $x$ increases. 
  Note that here, $f(x,\Delta)$ is singular at $\Delta = 0$ 
  provided ${\bar x} < \frac{1}{8}$ 
  (see~\eq\nr{small-Delta-condition}).
  } 
\la{fig:exact-quenching}
\end{figure}
%

Thermal fluctuations render the incident particle's damping rate 
in \eq\nr{damping-rate} formally divergent in the HTL framework, 
\be
\la{total-rate-med} 
  \Rate  
  \; = \;  
  \int_{-\infty}^{\infty} 
  \dd \eps \, w(\eps) 
  \; = \;  
  \alpha_s \CF  
  \int_{-\infty}^{\infty} \dd \eps \, 
  [ 1 + \nB\big(\eps \big) ] \; \Sfunc^{ }_\rmi{full}(\eps) = \infty  \ ,
\ee
due to $\nB(\eps) \simeq T/\eps$ and 
$
  \Sfunc^{ }_\rmi{full}(\eps) 
  \simeq \Sfunc^{ }_\rmii{T}(\eps) 
  \simeq \frac13
$ 
at small $\eps$. 
As mentioned in sec.~\ref{sec:kin-eq}, 
this implies that the zero-scattering probability vanishes and 
that there is no $\delta(\Delta)$-term in $f(x,\Delta)$, 
in contrast with $T=0$. 
When $T > 0$ the distribution $f(x,\Delta)$ satisfies 
the same initial condition as for $T = 0$, 
namely, $f({x=0},\Delta) = \delta(\Delta)$, 
but this initial Dirac distribution is immediately washed out 
at any finite $x\,$, 
%
%
since its coefficient $e^{-x \Gamma}$ vanishes 
due to the divergence of $\Gamma$ at $T>0$ (see \eq\nr{total-rate-med}).
However, the $\delta(\Delta)$-term at $T=0$ is effectively replaced by 
a smeared distribution of the type \nr{eq:f-small-Delta}, 
contributing similarly to the normalization of $f(x,\Delta)$. 
The correspondence between the singular distribution \nr{eq:f-small-Delta} 
and a ``zero-scattering'' $\delta(\Delta)$-term at $T=0$ 
is also supported by the fact that \nr{eq:f-small-Delta} appears 
when $x$ is smaller than $(\alpha_s \CF T)^{-1}$
(see \eq\nr{small-Delta-condition}). 
Indeed, this scale is expected to be
the heavy quark mean free path (up to some logarithmic factor) 
in a thermal medium beyond HTL, 
due to screening of the transverse gluon propagator by the magnetic mass.

The above discussion helps explain why in spite of the drastic change 
between the $T = 0$ and $T\neq 0$ total scattering rates, 
the shape of $f(x,\Delta)$ changes continuously when progressively 
heating the medium. 
It also suggests that the divergence of $\Rate$ within 
the HTL framework should have a minor role in our calculation. 
This could be inferred from the distribution \nr{eq:bromwich} 
not depending explicitly on the total rate $\Gamma$, 
the integrals appearing in the expression \nr{eq:Inu-3} of $I(\nu)$ 
being regularised at $\eps \to 0$ by the presence of 
the Debye mass $\mD$ alone.

Finally, we consider the limit 
where $\mu \to 0$ and the hard scale is $\Lambda = \pi T$. 
In fig.~\ref{fig:exact-quenching} we show the collisional quenching weight 
$f(x,\Delta)$ obtained 
for $T = 3 m^{ }_\rmii{D}$ (corresponding to $g \approx 0.27$ for $\nf=3$),
and various values of $\bar x$. 
The curves may be directly compared with their $T=0$ counterparts 
in fig.~\ref{fig:T=0}. 
Here however, the displayed functions for $x>0$ are normalised 
and extend to $\Delta < 0\,$, 
exhibiting the feature of energy gain. 
We also note that all curves shown in fig.~\ref{fig:exact-quenching} 
correspond to the situation where 
$\frac{2}{3}x \alpha_s \CF T > 1$ (due to our choice of a quite large temperature),
except for $\bar x = 0.1$,
in which case the singularity at $\Delta = 0$ is clearly visible 
(right panel of fig.~\ref{fig:exact-quenching}).

%
\section{Discussion}
\label{se:disc}

We have calculated the collisional quenching weight $f(x,\Delta)$ in 
the limit $E^{ }_\ini \to \infty$ of a parton 
crossing a QCD medium characterised by the temperature $T$ 
and/or a quark chemical potential $\mu\,$. 
The main ingredient is the elastic scattering rate $w(\eps)$ 
in an elementary collision (defined by~\eq\nr{eq:w-eps-general}) 
obtained from a 
perturbative calculation combining HTL resummation for soft energy 
transfers and exact kinematics in the hard domain.\footnote{%
  In our study we are discussing a question of principle where 
  these energy scales are well-separated in the perturbative limit $g \ll 1$.
} 
In this setup, the rate ${\rm d}\Rate/{\rm d}\eps$ interpolating 
between the two regimes, defined by \eq\nr{rate-full}, 
should be valid from $\eps \sim \mD \sim g \Lambda$ 
(where $\Lambda \equiv {\rm max}(\pi T, \mu)$) up to arbitrarily large 
$\eps \gg \Lambda$, where the asymptotic behaviour 
in \eq\nr{eq:S_hard_expansions} holds. 
Our main results for $f(x,\Delta)$ are presented in fig.~\ref{fig:T=0} 
for a cold dense system, and in fig.~\ref{fig:exact-quenching} 
for a thermal medium. 

Within the HTL setup there is no magnetic screening, 
resulting in an infrared divergent total rate $\Rate$. 
However, as noted at the end of sec.~\ref{sec:S-soft}, 
$f(x,\Delta)$ does not depend explicitly on $\Rate$ and is infrared safe. 
Therefore, our results regarding $f(x,\Delta)$ are consistent 
at leading order 
and should be robust to ultrasoft corrections which would render 
$\Rate$ finite. 
In sec.~\ref{sec:exact-f}, an analogous situation was encountered, 
by starting from a cold dense medium (with finite $\Rate$) 
and slowly switching on temperature (making $\Rate$ infinite), 
see fig.~\ref{fig:small-T} and the related discussion. 
Figure~\ref{fig:small-T} gives some intuition on the impact of 
a finite versus infinite $\Rate$ on the shape of $f(x,\Delta)$. 

As recalled above, we have obtained $f(x,\Delta)$ in the 
$E^{ }_\ini \to \infty$ limit. 
Let us now discuss the effect of keeping $E^{ }_\ini$ finite on our results. 
Keeping $E^{ }_\ini$ finite introduces an upper cut-off 
$\eps^{ }_{\rm max}$ 
on the energy transfer $\eps$ in a single elastic collision. 
We thus expect $f(x,\Delta)$ obtained for $E^{ }_\ini \to \infty$ 
to be modified in those domains of $x$ and $\Delta$ which 
are sensitive to large $\eps\,$. 
As we have seen, large $\eps$ means small $\nu$, 
and $\nu \to 0$ is a sufficient condition for  
$f(x,\Delta)$ to approach the Landau distribution 
(see sec.~\ref{sec:Landau-limit} and eqs.~\nr{Landau-limit}-\nr{phiL-def}). 
From sec.~\ref{sec:Landau-limit}, 
one infers that the Landau distribution arises dominantly 
from $\eps$ and $\nu$ integration regions where 
$
  \eps 
  \, < \, 
  1/\nu \sim {\rm max}(\Delta, \, \mD\, \bar{x} \log{\bar{x}})
$. 
Thus, with a finite $E^{ }_\ini\,$, 
the distributions $f(x,\Delta)$ obtained for $E^{ }_\ini \to \infty$ 
remain valid as long as 
\be
  \la{eq:val-cond}
  {\rm max}(\Delta, \, \mD\, \bar{x} \log{\bar{x}}) 
  \; \ll  \;
  \eps^{ }_{\rm max} 
  \; \simeq \; 
  \frac{4 k E_\ini^2}{M_{ }^2 + 4 k E^{ }_\ini} 
  \; \simeq \;
  \left\{
  \begin{array}{l}
    \displaystyle \ E^{ }_\ini 
    \\[1mm]
    \displaystyle 4 k E^{ 2}_{\ini}/M^2
  \end{array} \right. 
  \ \mbox{for} \ 
  \left.
  \begin{array}{l}
    E^{ }_\ini \gg M_{ }^2/\Lambda  
    \\[1mm]
    E^{ }_\ini \ll M_{ }^2/\Lambda  
  \end{array} \right. 
  \; ,
\ee
where the expression of $\eps_{\rm max}$ holds for 
a target thermal parton of momentum $k \sim \Lambda$, 
and assuming $E_\ini \gg M \gg \Lambda$. 
In what follows, we use this upper cutoff only to estimate 
the finite-$E_\ini$ effects in the regime where 
the fractional energy loss remains negligible. 

When $\eps^{ }_\rmi{max}$ is kept finite, 
the average energy loss 
$
  \langle \Delta \rangle 
  \equiv  
  \int {\rm d} \Delta\, f(x,\Delta) \, \Delta 
$ 
becomes well-defined (which is not the case in 
the $E_\ini \to \infty$ limit, where $f(x,\Delta) \sim 1/\Delta^2$ at 
large $\Delta$). Returning to the kinetic equation~\nr{eq:kinetic1}, 
we may now write $\eps^{ }_\rmi{max}$ for the upper limit 
of integration because $w(\eps) = 0$ 
for $\eps > \eps^{ }_\rmi{max}\,$. 
(As for the lower limit of integration, 
as in our study we can take $-\infty$, recalling that the contribution 
from energy gain, $\eps < 0$, 
is exponentially suppressed when $-\eps \gg T\,$.) 
%
%
Multiplying \eq\nr{eq:kinetic1} by $\Delta$ and integrating over $\Delta$, 
we easily see that $\langle \Delta \rangle$ is, as expected, linear in $x$, 
with a coefficient given by the first moment of $w(\eps)$:
\be
  \la{eq:ave-loss}
  \langle \Delta \rangle 
  \; = \; 
  x  
  \int^{\eps^{ }_\rmii{max}}_{-\infty} {\rm d} \eps\, 
  w(\eps) \, \eps 
  \; = \;  
  x\, \alpha_s \CF 
  \int_{0}^{\eps_\rmii{max}} {\rm d}\eps\, 
  \Sfunc(\eps) \, \eps  \ , 
\ee
where $\Sfunc$ is defined by~\eq\nr{eq:w-eps-general} and 
the contribution from energy gain is expressed in terms 
of $\eps > 0$ as in \eq\nr{eq:Inu-2}. 
Then, recalling 
\eq\nr{eq:S_hard_expansions}, 
we see that there are two 
dominant logarithmic intervals contributing to $\langle \Delta \rangle $ 
in the range between $\mD$ and $\eps^{ }_\rmi{max}\,$, 
namely $\mD \ll \eps \ll \Lambda$ 
and $\Lambda \ll \eps \ll \eps^{ }_\rmi{max}\,$, 
giving to logarithmic accuracy: 
\be
\la{eq:Bjorken-loss-generalized}
  \langle \Delta \rangle
  \; \simeq \; 
  x \alpha_s \CF \bigg\{
    \int_{\mD}^\Lambda {\rm d}\eps \,
    \frac{\mD^2}{2\eps}
    +
    \int_{\Lambda}^{\eps_\rmii{max}} {\rm d}\eps \, 
    \frac{\mD^2}{4\eps}
  \bigg\} 
  \; = \;
  x \alpha_s \CF \frac{\mD^2}{4} 
  \log \frac{\Lambda \eps_\rmi{max}}{\mD^2} 
  \; .
\ee
In the case of zero chemical potential ($\Lambda \sim T$) 
and for $E_\ini \gg  M^2/T$ (implying $\eps_\rmi{max} \simeq E_\ini$), 
we recover Bjorken's leading-log result for the average loss 
of a light quark in a QGP~\cite{Bjorken:1982tu}. 

The average loss \nr{eq:Bjorken-loss-generalized} arises 
from a $\Delta$-integration 
domain where $\Delta \ll \eps_\rmi{max}\,$. 
When $\eps_{\rm max} \gg \mD\,\bar{x}\log{\bar{x}}\,$, 
the distributions $f(x,\Delta)$ obtained for $E^{ }_\ini \to \infty$ are valid as long as 
$\Delta \ll \eps_\rmi{max}$ (see \eq\nr{eq:val-cond}). 
%
%
We conclude that all $f(x,\Delta)$ in our study have a 
regulated average, obtained once 
the finite physical cutoff $\eps_\rmi{max}$ is imposed, 
which coincides with the leading-log result \nr{eq:Bjorken-loss-generalized}. 
This is so for any $x$, even though the shape of $f(x,\Delta)$ 
is clearly quite different at different values of $x\,$. 
Note that when $E_\ini$ is sufficiently large at fixed $x$, 
the average $\langle \Delta \rangle$ scales 
logarithmically with $\eps^{ }_\rmi{max} \simeq E_\ini$ 
and is much larger than 
the {\em most probable} $\Delta\,$, defined as the position $\Delta^{ }_p$
of the peak of $f(x,\Delta)$, which remains constant in the $E_\ini \to \infty$ limit. 
At large (but fixed) $x$, the most probable $\Delta$ follows from the 
Landau scaling behaviour~\nr{Landau-limit}. Noting that the function
$\varphi \left( \lambda \right)$ reaches its maximum at $\lambda \approx -0.22\,$, we obtain
\be
\la{eq:Delta-p}
  \Delta^{ }_p 
  \; \simeq \; 
  x \alpha_s \CF \frac{\mD^2}{4} \,  
  \big( \log \bar x + \kappa - 0.22 \big) 
  \; \simeq \;
  x \alpha_s \CF \frac{\mD^2}{4} 
  \log \frac{\Lambda \bar x}{\mD}  \; ,
\ee
where the second approximation follows from 
$\kappa \simeq \log \frac{\Lambda}{\mD}$, which holds 
in the perturbative limit ($\mD \ll \Lambda$), and to 
logarithmic accuracy.\footnote{%
  $\kappa \simeq \log \frac{\Lambda}{\mD}$ is easily obtained by setting $\Lambda^\star = \mD$ 
  in eqs.~\nr{I1-small-nu} and \nr{a1a2-def} 
  ($\kappa$ being independent of $\Lambda^\star$). 
  In view of comparing $\langle \Delta \rangle$ and $\Delta^{ }_p$ below, 
  we need to keep this logarithmic dependence of $\kappa$ for consistency, 
  since the same type of logarithm is kept 
  in \eq\nr{eq:Bjorken-loss-generalized}.
}

At sufficiently large $E_\ini\,$, the average $\Delta$ arises mostly from 
a rare but hard single scattering, rather than from the accumulation of soft 
interactions. This is why the central limit theorem breaks down and the distributions 
we find are markedly non-Gaussian.  
Decreasing $E^{ }_\ini$ would cause $\eps^{ }_\rmi{max}$ 
to change according to \eq\nr{eq:val-cond} and to eventually decrease 
faster than $E^{ }_\ini$ as soon as $E^{ }_\ini < M^2/\Lambda$.
When $E_\ini$ further decreases, $\langle \Delta \rangle$ and $\Delta^{ }_p$ 
become of the same order when $\eps_{\rm max} \sim \mD\, \bar{x}$ 
(see eqs.~\nr{eq:Bjorken-loss-generalized} and \nr{eq:Delta-p}), 
which is precisely the scale where the distribution $f(x,\Delta)$ 
obtained for $E^{ }_\ini \to \infty$ starts to be invalid. 
As found by Vavilov~\cite{Vavilov:1957zz}, 
when $\eps_{\rm max} \ll \mD\, \bar{x} \, \ll  E_\ini\,$, the distribution approaches
a Gaussian (whereby $\Delta^{ }_p = \langle \Delta \rangle\,$, 
with $\langle \Delta \rangle$ given by~\eq\nr{eq:Bjorken-loss-generalized} 
in terms of $\eps_{\rm max}$). This is expected since this situation formally corresponds to 
the limit of large $x$ at fixed $\eps^{ }_\rmi{max}$, 
and is thus compatible with the central limit theorem.

The discussion above highlights that the structure of $f(x,\Delta)\,$, 
obtained here from a perturbative calculation, 
depends sensitively on the interplay between the 
scales $x$ and $E^{ }_\ini\,$, and notably 
on the order in which the limits $E^{ }_\ini \to \infty$ 
and $x \to \infty$ are taken. 
Our work provides the numerical solution 
for the collisional quenching weight $f$ at arbitrary path lengths $x\,$, 
including both the Landau limit 
(for large $x$, sec.~\ref{sec:Landau-limit}) 
and the finite-$x$ regime, where fluctuations 
and energy gain play a prominent role. 
The distributions we find exhibit rich non-Gaussian behaviour, \eg \ they are 
dominated by rare hard scatterings at large $\Delta\,$, and also have 
a singularity at $\Delta = 0$ for sufficiently small $x$ (see \eq\nr{eq:f-small-Delta} 
and figs.~\ref{fig:small-T} and \ref{fig:exact-quenching}).
While a detailed phenomenological study is beyond the 
scope of the present work, 
it is tempting to speculate that such effects may 
be relevant in situations where the path length $x$ is limited 
and fluctuations are not efficiently averaged out. 
In particular, a probabilistic treatment of collisional energy loss 
(including the aspect of energy gain) 
may provide useful insight 
in light-ion or high-multiplicity small systems, 
where the applicability of standard quenching 
frameworks is still under active investigation.

%
\section*{Acknowledgements}

We thank Jacopo Ghiglieri and Andr\'e Peshier for useful feedback on our work.
G.J.\ is funded by the Agence Nationale de la Recherche (France), 
under grant ANR-22-CE31-0018. 

\vspace{3mm}
%
\section*{Note added}

The recent ref.~\cite{DuPlessis:2026pyr} considered heavy quark 
collisional energy loss beyond leading-log accuracy, 
but worked explicitly in the late-time limit from the outset 
(corresponding to $x \to \infty$ for us) with $E_\ini$ finite. 
They also found non-Gaussian tails, but of a 
totally different nature (being exponential and related 
to detailed balance). 
On the contrary, we focus 
on the case of a {\em tagged} heavy quark 
crossing a plasma %
(allowing one to take the $E_\ini \to \infty$ limit), 
so that thermal equilibration is beyond the scope of our study.

%
\appendix
\renewcommand{\thesection}{\Alph{section}} 
\renewcommand{\thesubsection}{\Alph{section}.\arabic{subsection}}
\renewcommand{\theequation}{\Alph{section}.\arabic{equation}}

%
\section{Scattering rate from gluon spectral function}
\label{app:xsection}

We consider a projectile heavy quark (mass $M$) crossing 
a QCD medium in thermal equilibrium, 
of temperature $T$ and chemical potential $\mu$, 
characterised by the ``hard scale'' $\Lambda \equiv {\rm max}(\pi T, \mu)\,$. 
We assume $M \gg \Lambda$ (allowing the heavy quark to be 
unambiguously tagged when exiting the medium) 
and a heavy quark velocity $v \sim \rmO(1)$ 
(but not necessarily $v \to 1$ in this appendix). 

In the context of energy loss, 
the heavy quark Fock space involves the transition between 
the single particle states of momenta $\vec{p}_\ini $ and $ \vec{p}_\fin\,$, 
namely, $| \vec{p}_\ini \rangle = \hat a^\dagger_{\vec{p}_\ini} | 0 \rangle$ 
and $| \vec{p}_\fin \rangle = \hat a^\dagger_{\vec{p}_\fin} | 0 \rangle$, 
the state $|0 \rangle$ referring to the heavy quark vacuum. 
The interaction Hamiltonian is given by
\be
 \hat{H}^{ }_\rmii{I}(t) 
 \; = \; - i g \int \! \dd^3 \vec{x}\ 
 \bar{\psi}_{i} (\X) \gamma^\mu \psi_{j} (\X)
 \,
 t^a_{ij}
 A^a_\mu (\X)
 \;, \hspace*{5mm}
 \la{H_I_full}
\ee
where $\X \equiv (t,\vec{x})$, 
$i$ and $j$ are fundamental colour indices, 
$a$ is an adjoint colour index, 
and $g$ is the strong coupling. 
Note that $\hat{H}^{ }_\rmii{I}(t)$ describes 
the interaction between the heavy quark field and the thermal ensemble, 
but the coupling $g$ also describes interactions among 
the thermal degrees of freedom. 

When studying the energy exchange between the plasma 
and a fast ``test particle'', 
the integrated interaction rate is not sufficient. 
One needs double-differential scattering rates which, 
for a transition $\vec{p}_\ini \to \vec{p}_\fin$ involving 
a single gluon exchange in the $t$-channel, 
can be obtained as follows~\cite{Jackson:2024gtr}. 

We consider initial and final states of the form
\be
 |I\,\rangle \equiv 
 |\ini \, \rangle \otimes 
 | \vec{p}_\ini \,\rangle 
 \;, \quad
 |F\,\rangle \equiv 
 |\fin \,\rangle \otimes 
 | \vec{p}_\fin \,\rangle
 \;, \la{states_scat}
\ee
where $|\ini\,\rangle$ and $|\fin\,\rangle$ are 
the initial and final states of the QCD medium. 
The set of possible states $|\ini\,\rangle$ 
(which coincides with the set of states $|\fin\,\rangle$) 
is a basis of thermal states. 
The states $|\ini\,\rangle$ can be chosen as eigenvectors of 
both the Hamiltonian $\hat H$ and the operator $\hat N$ associated 
with quark number (since these operators are Hermitian and commute), 
with eigenvalues denoted by ${\cal E}^{ }_{\sini}$ and ${\cal N}^{ }_{\sini}$, 
respectively.

To first order in the interaction Hamiltonian, 
the transition matrix element reads
\be
 T^{ }_\rmi{$F$$I$} =
 \langle F\, | \int_0^{t} \! {\rm d}t' \, \hat H_\rmii{I}^{ }(t') \, 
 | I\,\rangle
 \;. \la{transition_matrix}
\ee
A quantum-mechanical version of the transition rate is 
then given by Fermi's golden rule, 
\be
 \Theta 
 \big( \vec{p}_\ini \to \vec{p}_\fin \big)
 \; \equiv \; 
 \lim_{t,V\to \infty} 
 \sum_{\sini,\,\sfin}
 \frac{e^{
    - \beta ({\cal E}^{ }_{\sini} - \mu \, {\cal N}^{ }_{\sini} )
  }}{{\cal Z}^{ }_\rmiii{QCD}}
 \frac{| T^{ }_\rmii{$F$$I$}|^2_{ }}{t\, V}
 \;, \la{def_Theta} 
\ee
where $\beta \equiv 1/T$, 
and the thermal partition function reads 
$
  {\cal Z}^{ }_\rmii{QCD} 
  \equiv 
  \sum_{\sini} 
  e^{- \beta ({\cal E}^{ }_{\sini} - \mu \, {\cal N}^{ }_{\sini} )}
$. 
Inserting the plane wave expansions of $\psi_{j}$ and $\bar{\psi}_{i}$ 
(in terms of creation and annihilation operators) 
into the Hamiltonian \nr{H_I_full}, we obtain
\be
\la{TFI-expanded}
 T^{ }_\rmi{$F$$I$} \, =  \, 
 g \, \frac{
 \bar{u}^{ }({\vec{p}_\fin},\sigma_\fin)
 \gamma^{\mu} 
 {u}^{ }({\vec{p}_{\ini}},\sigma_\ini)
           }
      { \sqrt{ (2 E^{ }_\ini \, )(2 E^{ }_\fin\,) } }
 \,
 \int_{\X'}
 \,
 \langle \fin \, | t_{ij}^a A^a_\mu(\X') | \ini \, \rangle
 \; 
 e^{i (\P^{ }_\ini\, - \P^{ }_\fin\,)\cdot \X' }_{ }
 \; , 
\ee
where $\int_{\X} \equiv \int \! \dd t \int \! \dd^3 \vec{x}$. 
Inserting now \nr{TFI-expanded} in \eq\nr{def_Theta} and 
then using $\tr( t^a t^b ) = \frac12 \delta^{ab}$, 
spacetime translational invariance, 
the completeness property 
$\sum_{\sfin}| \fin\,\rangle \langle \fin \, | = \unit$, 
and the notation 
$
  \langle...\rangle^{ }  
  \equiv  
  \sum_{\sini} 
  e^{- \beta ({\cal E}^{ }_{\sini} - \mu \, {\cal N}^{ }_{\sini} )} 
  \langle \ini \,|... | \ini\,\rangle / {\cal Z}^{ }_\rmii{QCD}
$ for thermal expectation values, we arrive at
\be
 \Theta 
 \big( \vec{p}_\ini \to \vec{p}_\fin \big)
 \; = \; 
 \frac{ g^2 \delta^{ab}
 \, L^{\mu\nu}(\P_\fin,\P_\ini)
        }{8 E_\fin\, E_\ini}
 \int_{\X}
 \bigl\langle 
   {A}_{\mu}^{a}(\X) \, 
   {A}_{\nu}^{b}(0) 
 \bigr\rangle
 \, e^{i\Q\cdot\X }_{ }
 \; , \la{wightman_1}
\ee
where $\Q \equiv \P_\ini - \P_\fin \equiv (\eps, \vec{q})$ 
is the exchanged four-momentum, 
and $L^{\mu\nu}$ the leptonic trace originating from 
the sum over the initial and final heavy quark helicities:
\bea
  {\sum_{\sigma_\fin,\sigma_\ini}}\;
  \bar{u}^{ }({\vec{p}_\fin},\sigma_\fin)
  \gamma^{\mu} 
  {u}^{ }({\vec{p}_{\ini}},\sigma_\ini)
  \, 
  \bar{u}^{ }({\vec{p}_\ini},\sigma_\ini)
  \gamma^{\nu} 
  {u}^{ }({\vec{p}_{\fin}},\sigma_\fin)
  \, = \,  
  \tr \bigl[\, 
  ({\bsl\P}^{ }_{\!\!\fin}+M)
  \gamma^{\mu} 
  ({\bsl\P}^{ }_{\!\!\ini}+M)
  \gamma^{\nu} 
  \,\bigr] \hskip 10mm && 
  \nn[-1mm]
  \,  = \, 
  4 \bigl(\;
  \P_\fin^\mu \P_\ini^\nu +
  \P_\ini^\mu \P_\fin^\nu +
  g_{ }^{\mu\nu} 
  \, (M^2 - \P^{ }_\ini \cdot \P^{ }_\fin)
  \; \bigr) 
  \, \equiv \, 
  L^{\mu\nu}(\P^{ }_\fin \,,\P^{ }_\ini) 
  \; .  && 
  \la{L-def}
\eea

The rate given in \eq\nr{wightman_1}, 
expressed in terms of a thermal correlation function of gauge fields, 
is a fundamental quantity from which many observables can be derived. 
For instance, if one is interested in studying, 
via Boltzmann equations, the equilibration of heavy quarks, 
which can proceed via elastic scatterings such as 
annihilation, Compton, 
and $t$-channel scattering (at small $t$), 
the ``coefficient'' $\Theta ( \vec{p}_\ini \to \vec{p}_\fin )$ 
enters that part of the collisional operator corresponding 
to the latter (see eq.~(2.19) of ref.~\cite{Jackson:2024gtr}). 
In our specific problem where the heavy quark is an incoming test particle, 
the rate of elastic scatterings is given by 
\be
  \frac{ 
  {\rm d}\Rate 
  }{
    {\rm d} \eps \, {\rm d}^3 \vec{q}
  }
  \; = \;  
  \frac{1}{2 \Nc}
  \int \! \frac{ \dd^3 \vec{p}_\fin }{(2\pi)^3}
  \ 
 \Theta 
 \big( \vec{p}_\ini \to \vec{p}_\fin \big)
 \; 
 \delta^{(4)} \big( \Q + \P_\fin - \P_\ini \big) \ , 
 \la{eq:gamma_el_0}
\ee
where the factor $\frac{1}{2 \Nc}$ accounts for the {\em average} over 
the incoming heavy quark spin and colour indices. 
(The latter indices are simply summed over in the derivation 
of \nr{wightman_1}, as needed, for example, 
in the heavy-quark equilibration problem.) 

The two-point correlator in \eq\nr{wightman_1} is 
a Wightman function, which can be written as~\cite{Laine:2016hma}
\be
 \int_{\X}
 \bigl\langle
   {A}_{\mu}^{a}(\X) \, 
   {A}_{\nu}^{b}(0) 
 \bigr\rangle
 \, e^{i\Q\cdot\X }_{ }
 \; = \; -  2 \pi \, \big[ 1 + \nB\big(\Q \cdot u \big) \big] 
 \, \delta^{ab} \, \rho_{\mu\nu}^{ }(\Q) \ ,
 \la{wightman_2}
\ee
with $u$ the medium's four-velocity, 
and $\rho_{\mu\nu}^{ }(\Q)$ the spectral function defined by\,\footnote{%
  The spectral function used in our study differs by a factor $\pi$ 
  from that of Ref.~\cite{Laine:2016hma}, hence 
  the overall factor $\pi$ in \eq\nr{wightman_2}, 
  and $\frac{1}{\pi}$ in eqs.~\nr{rhomunu-def-1} and \nr{rhomunu-def-2}.
} 
\be
\la{rhomunu-def-1}
  \delta^{ab} \,
  \rho_{\mu\nu}^{ }(\Q) 
  \; = \;  
  \frac{1}{2 \pi}
  \int_{\X} 
  \bigl\langle \big[ {A}_{\mu}^{a}(\X) , {A}_{\nu}^{b}(0) \big] \bigr\rangle 
  \, e^{i\Q\cdot\X }_{ }  \ . 
 \ee
Note that $\rho_{\mu\nu}^{ }(\Q)$ being a fortiori symmetric in 
$\mu \leftrightarrow \nu$, 
the property $\rho_{\mu\nu}^{ }(-\Q) = - \rho_{\mu\nu}^{ }(\Q)$ 
directly follows from \eq\nr{rhomunu-def-1} 
by changing the integration variable as $\X \to - \X$ 
and using translational invariance of the thermal averages.
 
As is well-known, the spectral density $\rho_{\mu\nu}^{ }(\Q)$ 
can be expressed in terms of the imaginary part of 
the gluon propagator as follows\,\footnote{%
  The gluon propagator $\Delta_{\mu\nu}^{ }(\Q)$ in Minkowski 
  space (with $\Q \equiv (\eps,\vec{q})$) is obtained from 
  the propagator $\Delta_{\mu\nu}^{ }(Q)$ computed in the 
  imaginary-time formalism (with $Q \equiv (q^{ }_n ,\vec{q})$, 
  where $q^{ }_n = 2\pi n \,T$ are bosonic Matsubara frequencies), 
  and then analytically continued using the prescription 
  $iq^{ }_n \to \eps + i 0^+$~\cite{Laine:2016hma}. 
  Note the Minkowski and Euclidean metric conventions: 
  $\Q^2 \equiv \eps^2 - q^2$ and $Q^2 \equiv q_n^2 + q^2\,$.
}
\be
  \la{rhomunu-def-2}
  \rho_{\mu\nu}^{ }(\Q) 
  \; = \;  
  \frac{1}{\pi} \im \Delta_{\mu\nu}^{ }(\eps+ i 0^+, \vec{q})  \ . 
\ee
In covariant gauge, the gluon propagator is 
\be
  \la{prop-cov-gauge}
  \Delta^{\mu\nu}( \Q )  
  \; = \; 
  \frac{ \mathbbm{P}^{\mu\nu}_\rmii{T}}{-\Q^2 + \Pi_\rmii{T}(\Q)}
  + \frac{ \mathbbm{P}^{\mu\nu}_\rmii{L}}{-\Q^2 + \Pi_\rmii{L}(\Q)}
  + \xi \frac{ \Q_{ }^\mu \Q_{ }^\nu }{\Q^4}  \ ,
\ee
where $\xi$ is the gauge parameter, 
$\Pi_\rmii{T,L}(\Q)$ are the transverse and longitudinal parts of 
the gluon polarisation tensor, and the projectors 
$\mathbbm{P}^{\mu\nu}_\rmii{T,L}$ are, 
for an isotropic medium, both orthogonal to $\Q^\mu$ and 
can be expressed in terms of $\Q^\mu$, $u^\mu$ and $g^{\mu \nu}$ 
(see ref.~\cite{Weldon:1982aq}). 

Using eqs.~\nr{wightman_2}, \nr{rhomunu-def-2} and \nr{prop-cov-gauge} 
in \eq\nr{wightman_1}, we see that the gauge dependent piece 
$\sim \Q_\mu \Q_\nu$ in \nr{prop-cov-gauge} does not contribute 
(since the tensor $L^{ }_{\mu\nu}$ defined by \nr{L-def} 
is orthogonal to $\Q_\mu$), and the transition rate can be rewritten as
\be
  \Theta  \big( \vec{p}_\ini \to \vec{p}_\fin \big) 
  \; = \; 
  - \frac{ (\Nc^2 -1) \pi g^2}{4 E_\fin\, E_\ini} \, 
  [ 1 + \nB\big(\Q \cdot u \big) ] \, 
  L_{\mu\nu}(\P_\fin,\P_\ini) \, 
  \big[  \mathbbm{P}^{\mu\nu}_\rmii{T} \, \rho_\rmii{T} + 
  \mathbbm{P}^{\mu\nu}_\rmii{L} \, \rho_\rmii{L} \big] \; , 
  \la{wightman_3}
\ee
where the transverse and longitudinal gluon spectral functions are defined by 
eqs.~\nr{propagators-0} and \nr{cut-0}. 
Note that these scalar functions can only depend on 
the Lorentz scalars $\Q \cdot u$ and $\Q^2$, and that 
\eq\nr{wightman_3} can in principle be used in any Lorentz frame.

Choosing now the medium's rest frame, $u=(1,\vec{0})$, 
we can trade $\Q \cdot u$ and $\Q^2$ for $\eps$ and $q = |\vec{q}|$, 
\ie\ $\rho_\rmii{T,L} \rightarrow \rho_\rmii{T,L}^{ }(\eps,q)$,\footnote{%
  Moving away from the rest frame by introducing a flow velocity $u^\mu$, 
  $\eps$ should be replaced by $\Q \cdot u$ and 
  $q$ by $\sqrt{ (\Q\cdot u)^2 - \Q^2}$.
}
and the projectors $\mathbbm{P}^{\mu\nu}_\rmii{T,L}$ obtained 
in ref.~\cite{Weldon:1982aq} simplify to
\be
 \mathbbm{P}^{\mu\nu}_\rmii{T} 
 \; \equiv \; 
 - \, g^{\mu}_{\  i} \, g^{\nu}_{\ j}
   \,\biggl( \delta_{ij} - \frac{q_i q_j}{q^2} \biggr)
 \;, 
 \quad
 \mathbbm{P}^{\mu\nu}_\rmii{L} 
 \; \equiv \; 
 g^{\mu\nu}
 - \frac{\Q^\mu \Q^\nu}{\Q^2}
 - \mathbbm{P}^{\mu\nu}_\rmii{T}
 \;. \la{projectors}
\ee
These projectors need to be contracted with $L^{\mu\nu}$ 
in \eq\nr{wightman_3}. 
We obtain:
\be
 L_{\mu\nu} 
 \mathbbm{P}^{\mu\nu}_\rmii{T} 
 \; \; = 
 - 8 \Bigl[  \, p_\ini^2 - \frac{(\vec{p}_\ini \cdot \vec{q})^2}{q^2} 
 - \frac{t}{2} \, \Bigr] \; ; 
 \hspace{8mm}
  L_{\mu\nu} 
 \mathbbm{P}^{\mu\nu}_\rmii{L} 
 \; = \; 
 8 \Bigl[\, 
 E_\ini^2 - \frac{(\vec{p}_\ini \cdot \vec{q})^2}{q^2} 
 \,\Bigr] \; ,
 \la{eq:contractions}
\ee
where the Mandelstam variable $t$ is defined by
\be
  t 
  \; \equiv \; 
  \Q^2 
  \; = \; 
  (\P^{ }_\ini - \P^{ }_\fin)^2 
  \; = \; 
  2 (M^2 - \P^{ }_\ini \cdot \P^{ }_\fin) 
  \; = \; 
  2\, \P^{ }_\ini \cdot \Q 
  \ . 
\ee

We then use \eq\nr{eq:contractions} and insert the transition 
rate \nr{wightman_3} in \eq\nr{eq:gamma_el_0}, 
where the integral over $\vec{p}_\fin$ is carried out trivially 
with the help of 
$\delta^{(3)} \big( \vec{q} + \vec{p}_\fin - \vec{p}_\ini \big)$. 
For the remaining energy conserving delta function, we use the identity
\be
  \delta \big( \eps - E_\ini + E_\fin \big) 
  \; = \; 
  2 E_\fin \ 
  \delta \big( (\eps - E_\ini )^2 - E_\fin^2 \big)
  \; = \; 
  \frac{E_\fin}{E_\ini} \, 
  \delta \bigg( \eps -  v q_{\parallel} - \frac{t}{2 E_\ini} \bigg) \ , 
\la{exact-delta}
\ee
where $v = p_\ini / E_\ini$ is the heavy quark velocity, 
and $q_{\parallel}$ the component of $\vec{q}$ parallel to $\vec{p}_\ini\,$. 
As a result, the differential rate \nr{eq:gamma_el_0} reads (with $\CF = (\Nc^2-1)/(2\Nc)$)
\bea
  \frac{ 
    {\rm d}\Rate 
  }{
    {\rm d} \eps \, {\rm d}^3 \vec{q}
  } 
  & = & 
  \frac{g^2 \CF}{(2\pi)^2} \left[ 1 + \nB\big(\eps\big) \right] 
  \, \delta \bigg( \eps -  v q_{\parallel} - \frac{t}{2 E_\ini} \bigg) 
  \nn[2mm]
  & \times & 
  \left\{ 
    \bigg[ v^2 - \frac{v^2 q_\parallel^2}{q^2} - \frac{t}{2 E_\ini^2}  \bigg] 
    \, \rho_\rmii{T}(\eps,q) 
    - 
    \bigg[ 1 - \frac{v^2 q_\parallel^2}{q^2} \bigg] 
    \, \rho_\rmii{L}(\eps,q) 
  \right\} \, .  
 \la{eq:gamma_el_exact}
\eea

We stress that no approximation was made to obtain 
\eq\nr{eq:gamma_el_exact} from \eq\nr{eq:gamma_el_0}, 
so that \eq\nr{eq:gamma_el_exact} is an exact expression 
of the differential rate associated to a single $t$-channel scattering, 
which, if necessary, can be used up to 
$|t| = |t|_{\rm max} = (s-M^2)^2/s\,$, where $s$ is 
the Mandelstam variable of the elastic scattering~(such hard exchanges
resolve the $2 \to 2$ scattering on thermal partons). 
Note that within the assumptions $M \gg \Lambda$ and $v \sim \rmO(1)$, 
the term $\sim {t}/{E_\ini^2}$ in the factor multiplying $\rho_\rmii{T}$ 
in \eq\nr{eq:gamma_el_exact} can always be neglected. 
This follows from writing 
$
  s = (\P^{ }_\ini + \K^{ }_\ini)^2 
  = M^2 + 2 \P^{ }_\ini \cdot \K^{ }_\ini\,
$ 
(where $\K^{ }_\ini = (k_\ini , \vec{k}^{ }_\ini)$ 
denotes the initial momentum of the thermal particle participating 
in the elastic scattering), and using
\be
  |t|_{\rm max} 
  \; \simeq \; 
  \frac{4 (1+v)^2 k_\ini^2 E_\ini^2}{M^2 + 2(1+v) k_\ini E_\ini} \ \ 
  \Longrightarrow \ \ \frac{|t|}{E_\ini^2} 
  \; \leq \;  \frac{|t|_{\rm max}}{E_\ini^2} 
  \; \ll \; 1 
  \ . 
\ee

Using \eq\nr{eq:gamma_el_exact}, 
we can recover the expression of the total rate $\Rate$ 
found in ref.~\cite{Peigne:2007sd} (in the context of QED), 
see eq.~(13) of this reference.\footnote{\la{foot:spectral-convention}%
  Let us note that there are different definitions of 
  $\rho^{ }_{\rmii{L}}$ in the literature. 
  The spectral density $\rho^{ }_{\rmii{L}}$ used in our study 
  is related to those of 
  refs.~\cite{Thoma:1990fm} and~\cite{Peigne:2007sd}, 
  denoted by $\rho_{\rmii{L}}^{\rmiii{[TG]}}$ 
  and $\rho_{\rmii{L}}^{\rmiii{[PP]}}$, 
  as 
  $
  \rho_{\rmii{L}}^{\rmiii{[TG]}} = - \rho_{\rmii{L}}^{\rmiii{[PP]}}  
  = - \frac{\Q^2}{q^2} \rho^{ }_{\rmii{L}}
  $.
} 
In ref.~\cite{Peigne:2007sd}, 
the quantity of interest is the average collisional loss beyond the 
leading logarithm, which requires controlling 
the exact kinematics of the elastic scattering up to transfers 
$|t| = |t|_{\rm max}$, 
and keeping the exact delta function in \nr{eq:gamma_el_exact}. 

Assuming $\eps/E^{ }_\ini \ll 1$ (in addition to $v \sim \rmO(1)$), 
it is easy to show that the delta function simplifies to 
$\delta \big( \eps -  v q^{ }_{\parallel} \big)$ 
(even if $\eps$ is not fixed when $E^{ }_\ini \to \infty$, 
\ie\ $\eps$ could still scale as $E^{ }_\ini\,$, 
provided $\eps/E^{ }_\ini \ll 1$). 
Starting from~\eq\nr{eq:gamma_el_exact}, 
one can then recover the expression used in ref.~\cite{Thoma:1990fm} 
to calculate the average loss to leading logarithmic accuracy.
In this study, for the sake of simplicity, 
we consider an {\em ultrarelativistic} heavy quark, 
and thus take $v \to 1$. 
The differential rate \nr{eq:gamma_el_exact} then
simplifies to the expression \nr{eq:gamma_el}, 
which serves as the starting point for our study. 

%
\section{
  Asymptotic expansions of 
  $\Sfunc_\rmii{T}(\eps)$ and $\Sfunc_\rmii{L}(\eps)$
}
\label{app:Taylor-ex}

In section~\ref{sec:S-soft}, we derived the functions $\Sfunc_i(\eps)$ 
(with $i=\{ \rmi{T}, \rmi{L} \}$), 
obtained from \eq\nr{eq:Sofeps} by using HTL gluon spectral densities. 
In this appendix, we collect the asymptotic expansions of these functions 
in three different limits: 
$\eps \ll \mD$, $\eps \to \frac{1}{\sqrt{3}}\,\mD$, and $\eps \gg \mD$. 

The expansions of $\Sfunc_i(\eps)$ follow from similar expansions for 
the solutions $q_{i,\eps}$ of the dispersion relations 
$\epsilon^2 = q^2 + \Pi_i (\epsilon, q)$, 
where $\Pi_i$ are the HTL self-energies \nr{eq:HTL_T}-\nr{eq:HTL_L}. 
(The first terms of the expansions of $q_{\rmii{T},\eps}$ 
and $q_{\rmii{L},\eps}$ were obtained in ref.~\cite{Weldon:1982aq}.) 
Inserting these into eqs.~\nr{ST-eps-HTL}--\nr{SL-eps-HTL} yields, 
for $\eps >0\,$:  
\bea
 \Sfunc^{ }_\rmii{T}(\eps)
&=& 
 \left\{
  \begin{array}{ll}
  \displaystyle 
   \!\! \frac13
   \; - \;
  \frac{8}{9}
  \biggl( 
  \frac{2\, \eps}{\pi^2\, \mD }
  \biggr)^{\frac{2}{3}}
  \!\! - \;
  \frac{160}{27}
  \biggl( 
  \frac{2\, \eps}{\pi^2\, \mD }
  \biggr)^{\frac{4}{3}}
  \!\! + \;
  \ldots 
  & 
  \mbox{for} \quad \eps \ll \mD^{ }\,,
  \\[4mm]
  \displaystyle 
   \!\! \frac{3}{28}
   - 
  \frac{2825}{10584 \sqrt{3}}
  \biggl(
  \frac{\eps}{\mD} - \frac1{\sqrt{3}}
  \biggr)
  + 
  \frac{269725}{1629936}
  \biggl(
  \frac{\eps}{\mD} - \frac1{\sqrt{3}}
  \biggr)^2
   + 
  \ldots
  & \displaystyle
  \mbox{for} \quad \eps \simeq \frac{\mD^{ }}{\sqrt{3}} \,,
  \\[4mm]
  \displaystyle
   \!\! \frac{ \mD^2 }{6\, \eps^2} 
  \;+\;  \frac{ \mD^4 }{72 \, \eps^4} 
  \biggl(
  55 - 18 \log \frac{8 \eps^2}{\mD^2} 
  \biggr) \;+\;
  \ldots
  & 
  \mbox{for} \quad \eps \gg \mD^{ } \,;
  \end{array}
 \right.
 \nn[-1mm]
 \la{ST_expansions} 
 \\[2mm]
 \Sfunc^{ }_\rmii{L}(\eps)
 &=& 
 \left\{
  \begin{array}{ll}
  \displaystyle 
  \!\! \frac{\pi \, \eps}{4 \, \mD }
  \;+\;
  \biggl( \frac{\pi^2}{4} - 4 \biggr)
  \frac{\eps^2}{ \mD^2 }
  \;+\;
  \frac{ 5 \pi}{128} 
  \big( 7 \pi^2 - 48 \big) 
  \frac{\eps^3}{ \mD^3 }
  \;+\;
  \ldots
  & 
  \mbox{for} \quad \eps \ll \mD^{ }\,,
  \\[4mm]
  \displaystyle 
  \!\! \frac{4}{21}
  + 
  \frac{200}{1323\sqrt{3}}
  \biggl(
  \frac{\eps}{\mD} - \frac1{\sqrt{3}}
  \biggr)
  -
  \frac{900}{3773}
  \biggl(
  \frac{\eps}{\mD} - \frac1{\sqrt{3}}
  \biggr)^2
  +\;
  \ldots
  \hspace*{6.5mm}
  & \displaystyle
  \mbox{for} \quad \eps \simeq \frac{\mD^{ }}{\sqrt{3}} \,,
  \\[4mm]
  \displaystyle
  \!\! \frac{ \mD^2 }{3\, \eps^2 - \mD^2}  
  \;+\;
  \ldots
  & 
  \mbox{for} \quad \eps \gg \mD^{ } \, .
  \end{array}
 \right. 
 \nn[-1mm]
 \la{SL_expansions} 
\eea
Note that in the expansion of 
$\Sfunc_\rmii{L}(\eps)$ at large $\eps \gg \mD$, 
the quoted term in  \eq\nr{SL_expansions} actually resums 
all powers of $\eps$, other corrections to 
$\Sfunc_\rmii{L}(\eps)$ in that limit being exponentially suppressed.
Let us also remark that since the functions $\Sfunc_i(\eps)$ are odd, 
expansions of these functions for $\eps <0$ are obtained by replacing 
$\eps \to |\eps|$ in eqs.~\nr{ST_expansions}-\nr{SL_expansions}, 
and multiplying by an overall minus sign. 
In the case of the function $\Sfunc_\rmii{T}(\eps)$, for 
which $\Sfunc_\rmii{T}(\eps \to 0^+) = \frac{1}{3}$, 
this leads to a discontinuity at $\eps =0$. 
See the end of sec.~\ref{sec:S-soft} for a discussion of this point.

%
\section{Method for numerical evaluation}
\la{app:numerics}

In this appendix we describe the method we have used 
to evaluate the solution to the kinetic equation $f(x, \Delta)$, defined by 
\eq\nr{eq:bromwich}, which involves integrating along a contour parallel 
to the imaginary axis (the Bromwich contour) in the complex $\nu$-plane. 
Since the expression for $I(\nu)$ in \eq\nr{eq:Inu-2} 
is only well-defined within the strip 
$0 \leq \re \nu \leq \beta \equiv \frac{1}{T}$,
the contour is in principle restricted to the fundamental strip 
as shown in fig.~\ref{fig:nu-contour}. 
We will choose the Bromwich contour to coincide with  
the imaginary axis, 
i.e. $\nu_0 = 0\,$. 
We thus need to evaluate $I(\nu)$ numerically 
for each $\nu$ lying on the imaginary axis. 
This procedure is nontrivial because 
the factor $e^{\nu \Delta}$ becomes highly oscillatory when $\nu$ is large, 
effectively turning the problem into 
the numerical Fourier transform of a complicated function (of $\nu$), 
with some accuracy for the desired value of $\Delta\,$.\footnote{%
  When the function of 
  $\nu$ can be analytically continued 
  without difficulty to the entire complex plane, 
  the Bromwich contour may 
  be deformed to a contour where the oscillations are damped. 
  One example is Talbot's contour~\cite{Talbot1979}. 
  However, in our case, there is no explicit formula for $I(\nu)$, 
  since the energy loss rate $w(\eps)$ is itself determined numerically. 
  Since it is not obvious how to analytically continue the function 
  to the entire complex plane, we choose to remain on 
  the imaginary axis to perform the integration over $\nu\,$.
}
Our numerical code 
as well as some data are available in ref.~\cite{zenodo_link}.

\begin{figure}[t]
\vspace*{1mm}
\centerline{
\includegraphics[scale=.6]{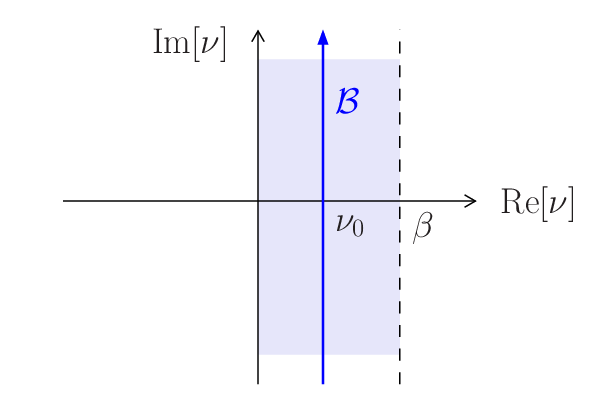}
}
\vspace*{-4mm}
\caption[a]{\small 
  The standard Bromwich contour (${\cal B}$) 
  used for taking the inverse Laplace transform, as 
  in \eq\nr{eq:bromwich}, needed to solve the kinetic equation determining $f(x,\Delta)\,$. 
  In our case, ${\cal B}$ 
  must fall within the strip of convergence 
  $0 \leq \re [ \nu ] \leq \beta \equiv 1/T$, 
  due to the integral defining $I(\nu)$. 
}
\la{fig:nu-contour}
\end{figure}
%

After parametrising the contour by $\nu \equiv i \eta / \mD$ 
where $\eta$ is a real variable, 
rescaling the energy $\epsilon \equiv \omega \mD$, 
and noting that the integrand is an even function of $\eta\,$, 
the expression \nr{eq:bromwich} of $f(x, \Delta)$ becomes 
(recalling \eq\nr{eq:w-eps-general}) 
\be
  \la{vac_sol1}
  f (x,\Delta)
  \; = \; \frac{1}{\mD} \int_0^\infty \frac{\dd \eta}{\pi} \,
  \cos \bigg(
  \eta\,
  \frac{ \Delta }{ \mD } 
  - \bar x \, H(\eta) 
  \bigg)
  \, 
  \exp \bigl\{ \,
  -\, \bar x \, G(\eta)
  \,
  \bigr\}  \; ,
\ee
where the functions $G$ and $H$ are given by 
\bea
  G(\eta) 
  & \equiv &
  4 \int_0^\infty {\rm d} \omega \,
  \Sfunc(\omega  \mD )
  \big[ 1+2\nB (\omega \mD ) \big]
  \big[ 
  1 - \cos(\omega  \eta) 
  \big]
  \la{eq:g}
  \; , \\
  H(\eta)
  & \equiv &
  4 \int_0^\infty {\rm d} \omega \,
  \Sfunc(\omega  \mD )
  \sin(\omega  \eta) 
  \la{eq:h}
  \; . 
\eea
Here we note that 
$G$ and $H$ also depend on the ratios $T/\mD$ and $\mu/\mD$ 
inherited from $\Sfunc = \Sfunc_\rmi{full}$ 
in \eq\nr{rate-full}. 
These functions are relatively straightforward 
to obtain numerically. 
However, in practice, it proves efficient to first tabulate $G$ and $H$ 
once (at some accuracy, for fixed $T$ and $\mu$) and then 
interpolate when performing the $\eta$-integration. 
%
%
Here we use  
$\eta^{ }_{\rm min} = 10^{-3}$ 
and 
$\eta^{ }_{\rm max} = 10^{3}$ 
with $\sim 10^4$ points 
(these values were chosen to ensure numerical convergence 
of all results plotted in our paper). 
Outside this range, it is necessary to use
the exact small and large $\eta$ behaviour 
for $G$ and $H$ for better accuracy. 
These may readily be derived using the asymptotic properties of $\Sfunc$ 
summarised in appendix~\ref{app:Taylor-ex}. 
In particular, as $\eta \to \infty\,$, 
we find $H \sim 4/(3\eta)$, 
while 
$
  G_0 
  \equiv 
  G(\eta)\big|_{\nB\to 0}
  \to 
  4 \int_{0}^\infty \! {\rm d}\omega \Sfunc(\omega \mD)\,
$, 
which is directly proportional to the total rate for a cold dense medium, 
cf. \eq\nr{total-rate-vac}. 
The term proportional to $\nB$ however 
has a different large-$\eta$ behaviour; 
it grows logarithmically, 
namely as 
$
  \Delta G(\eta) 
  \equiv G(\eta) - G^{ }_0(\eta)
  \sim \frac{8 T}{3 \mD} 
  \big( \log \eta + \gammaE \big)\,
$. 
This behaviour is shown in fig.~\ref{fig:h_and_g}. 

%
\begin{figure}[t]

\centerline{
\includegraphics[scale=.65]{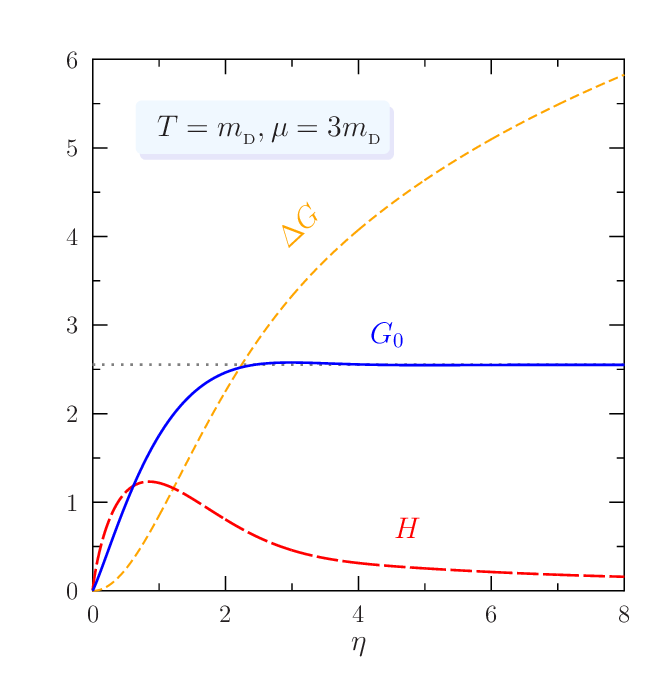}
}

\vspace*{-5mm}

\caption[a]{\small 
  The functions $G$ and $H$ 
  defined by eqs.~\nr{eq:g} and 
  \nr{eq:h} respectively. 
  Here $T=\mD$ and $\mu = 3 \mD$ (with $\nf=3$). 
  For the function $G$ we show $G_0 \equiv G(\eta)\big|_{\nB\to 0}$ 
  and $\Delta G \equiv G(\eta) - G_0(\eta)$ 
  to clearly separate the thermal effects 
  from the behaviour at $T=0\,$. 
} 
\la{fig:h_and_g}
\end{figure}
%

We note that if $T=0\,$, 
for large $\eta$ (other quantities being fixed)
the integrand in \eq\nr{vac_sol1} 
behaves approximately as $ \cos(\eta \Delta/\mD) e^{ - x \Rate }$, 
thus reproducing the $\delta$-component 
as discussed around \eq\nr{f-delta-piece}. 
This piece can be subtracted, so that although the integrand is rapidly 
oscillating (as a function of $\eta$) the modulus is damped 
and more amenable to numerical evaluation.\footnote{%
  To be concrete, we follow a method described in 
  sec.~5.3.2 of ref.~\cite{NumericalRecpies} for 
  oscillating integrals. 
  First, 
  the integral is split up into a sum 
  of smaller integrations  
  between successive zeros of the integrand, i.e. 
  \be
  \nonumber 
    {\cal I} 
    \; \equiv \; 
    \int_0^\infty {\rm d} \eta f(\eta) 
    \; = \; \sum_{i=0}^\infty {\cal I}_i \, ,
    \quad 
    {\cal I}_i = \int_{\eta_{i-1}}^{\eta_i} \dd \eta f(\eta)
    \, ,
  \ee
  where $f(\eta_i) = 0$ for $i = 0, 1, 2, ...$ and $\eta_{-1} = 0\,$. 
  Each individual term ${\cal I}_i$ can be evaluated relatively 
  simply by numerical integration, and then 
  the sum carried out over $i\,$.  
  This turns the problem into a series summation, for 
  which there are techniques enabling accelerations of convergence. 
  The zeros do not need to be exact but should only 
  be asymptotically correct.  
  We have used the so-called Levin $u$-transformation method 
  for accelerated convergence, 
  and chosen $\eta_i = (\frac12 + i)\pi \frac{\mD}{\Delta}\,$.
  }
For $T>0\,$, the tail of the $\eta$ integral 
is suppressed as $\eta^{- \frac23 x \alpha_s \CF T}$ 
as expected from \eq\nr{eq:Inu-large-nu-2}. 
We note that the frequency of the oscillations 
is controlled by $\Delta / \mD\,$.

%
\bibliographystyle{utphys}
{\small 
\bibliography{refs}
}

\end{document}